\documentclass[twocolumn,pre,aps,showpacs,floatfix]{revtex4}
\input{epsf}
\usepackage{graphicx}
\def \ind#1{{\mbox{\scriptsize {#1}}}}

\begin{document}

\title{Single Particle Jumps in a Binary Lennard-Jones System Below The
       Glass Transition}
\author{K. Vollmayr-Lee}
\email{kvollmay@bucknell.edu}
\affiliation{Department of Physics, Bucknell University,  
      Lewisburg, Pennsylvania 17837, USA}
\date{\today}

\begin{abstract}
We study a binary Lennard-Jones system below the glass transition with 
molecular dynamics simulations. To investigate the dynamics we focus
on events (``jumps'') where a particle escapes the cage formed by its
neighbors. Using single particle trajectories we define a jump by
comparing for each particle its fluctuations with  its changes in
average position. We find two kinds of jumps: ``reversible jumps,''
where a particle jumps back and forth between two or more average
positions, and ``irreversible jumps,'' where a particle does
not return to any of its former average positions, 
i.e. successfully escapes its cage. 
For all investigated temperatures both kinds of particles jump and
both irreversible and reversible jumps occur. With increasing temperature
relaxation is enhanced by an increasing number of jumps, and 
growing jump lengths in position and potential energy. 
However, the waiting time between two successive jumps is 
independent of temperature. This temperature independence might 
be due to aging, which is present in our system.
The times of jump duration are at all temperatures significantly shorter 
than the waiting times.
The ratio of irreversible to reversible jumps is
also increasing with increasing temperature, 
which we interpret as a consequence of the
increased likelihood of changes in the cages, i.e. a blocking of the
``entrance'' back into the previous cage. In accordance with this
interpretation, the fluctuations both in position and energy are
increasing with increasing temperature. 
A comparison of the fluctuations of jumping particles and 
non-jumping particles indicates that jumping particles are more mobile
even when not jumping.
The jumps in energy normalized by their fluctuations are decreasing
with increasing temperature, which is consistent with relaxation being
increasingly driven by thermal fluctuations. 
In accordance with subdiffusive behavior are the distributions of
waiting times and jump lengths in position.
\end{abstract}

\pacs{02.70.Ns, 05.20.-y, 61.20.Lc, 61.43.Fs, 64.70-Pf}
\maketitle

\section{Introduction}
\label{sec:intro} 
If a liquid is cooled and crystallization is avoided, one obtains a
Supercooled liquid. Upon further cooling the system falls
out of equilibrium and results in a glass. During the transition
from liquid to supercooled liquid to glass the thermodynamic
properties change and even more drastic changes occur in the
dynamics \cite{ref:glassintro}. The viscosity increases by many orders
of magnitude upon cooling 
and the mean square displacement (MSD) as a function
of time develops a plateau
at intermediate times. The lower the temperature the longer the
waiting time within the plateau until a second rise in the mean
square displacement occurs \cite{ref:MSDexamples}.
One common explanation for the plateau is that while at
high temperatures one has (at late enough times) normal diffusion,
at lower temperatures particles are 
caged in, i.e. trapped by their neighbors, and spend longer time 
within this cage with decreasing temperatures. The second increase in
the MSD indicates that after long enough waiting time the particles
manage to escape their cage. This escape out of the cage (``jump'')
is the focus of this paper.

To set the work of this paper in context, we review briefly 
previous studies on the dynamics of supercooled liquids
and glasses. We restrict ourselves mostly to simple
glass formers, excluding major work about strong glass formers, 
ionic systems, polymers and crystals.

Central quantities in both {\bf experiments and computer
simulations} are the viscosity, diffusion constant and MSD.
The MSD, which is an average over single particle $i=1,\ldots,N$ 
displacements, 
\begin{equation}
  r^2(t) = \frac{1}{N} \sum \limits_i | {\bf r}_i(t) - {\bf r}_i(0) |^2
  \hspace*{0mm}\mbox{,}
\end{equation}
and variations thereof \cite{ref:Oligschleger660-811} 
\begin{eqnarray}
\Delta R(t) & =  & ( N \cdot \mbox{MSD} )^{-1/2} \\
\Delta \overline{R(t)} & = & \left ( 
    \sum \limits_i \langle | {\bf r}_i(t) - {\bf r}_i(0) |^2 \rangle_{\Delta t}
                         \right )^{-1/2}\\
         & & \mbox{\small where } \langle \cdot \rangle_{\Delta t} \quad
                        \mbox{\small is a time average}  \nonumber \\
\Delta R^0(t) & = & \left (
       \sum \limits_i \left |{\bf r}^0_i(t) - {\bf r}^0_i(0) \right |^2
         \right )^{-1/2} \\
          & & \mbox{\small where } {\bf r}^0_i(t) \quad
           \mbox{\small is the position after steepest decent} \nonumber
\end{eqnarray}
show jumps when viewed with fine enough time resolution, and when
studied at low enough temperatures and for single configurations  
\cite{ref:Oligschleger660-811,ref:Teichler339-533,ref:Schober40,ref:Gaukel1907,ref:Caprion3709,ref:Schober723,ref:Sanyal4154,ref:Delaye2063}.
Detailed studies of these jumps indicate collective motion 
\cite{ref:Oligschleger660-811,ref:Teichler339-533,ref:Schober40,ref:Gaukel1907}.

One learns about the interplay between regular diffusion
and hopping motion via four-point correlation functions
\cite{ref:Lacevic030101-Glotzer509,ref:Glotzer214,ref:DoliwaA277,ref:Doliwa4915,ref:Roe1610} 
and the van Hove correlation function 
$G(r,t) = G_\ind{s}(r,t) + G_\ind{d}(r,t)$ 
\cite{ref:Roe1610,ref:Sanyal361,ref:Wahnstroem3752,ref:Bhattacharyya2741,ref:KlugePhD,ref:Gaukel1,ref:Gaukel664,ref:Schroeder331,ref:Roux7171,ref:Bruening964}
where $G_\ind{s}(r,t)$ is the self part and $G_\ind{d}(r,t)$ 
is the distinct part.
The hopping shows up in an additional peak in $G_\ind{s}(r,t)$ and 
an increase in the amplitude at $r \to 0$ of $G_\ind{d}(r,t)$.

Recently attention has been drawn to the non-Gaussian tail of $G(r,t)$
\cite{ref:Kob2827,ref:Donati2338-Donati3107,ref:Yamamoto4915,ref:Kegel290}
 and the non-Gaussian parameter $\alpha_2$
\cite{ref:Schober40,ref:DoliwaA277,ref:Kob2827,ref:Kegel290,ref:Weeks627,ref:Weeks095704,ref:Hurley10521,ref:Caprion4293,ref:RahmanA405,ref:Miyagawa8278}
which is the second coefficient of the cumulant series of 
$F_\ind{s}(q,t)$, the Fourier transform of $G_\ind{s}(r,t)$.
They are signatures 
of non-exponential behavior, which might 
be either due to homogeneous complex dynamics and/or due to
spatially heterogeneous dynamics \cite{ref:dynhetreviews}. 
This has been studied both with experiments
\cite{ref:Kegel290,ref:Weeks627,ref:Weeks095704,ref:exp_dynhet}
and in simulations of two-dimensional
\cite{ref:Hurley10521,ref:MuranakaR2735-Gould5707-Perera120-Doliwa32,ref:Hurley1694,ref:Perera5441,ref:Doliwa6898}
and three-dimensional systems
\cite{ref:Oligschleger660-811,ref:Schober40,ref:Gaukel1907,ref:Lacevic030101-Glotzer509,ref:Kob2827,ref:Donati2338-Donati3107,ref:Yamamoto4915,ref:Caprion4293,ref:Doliwa6898,ref:Gebremichael051503-Heuer6176-Yamamoto3515-Gaukel67-Poole51,ref:kvldynhetJCP,ref:Teichler717,ref:Laird636,ref:Schober6746,ref:Schober965}.
Studies of the most mobile particles reveal that they move
increasingly more collectively with decreasing
temperature. One distinct feature of this collective dynamics
is that some particles are moving along a string
\cite{ref:Oligschleger660-811,ref:Schober723,ref:Donati2338-Donati3107,ref:Weeks095704,ref:Schober965,ref:CaprionMRS2002,ref:Schober67}.

Similar motion has been found via the study of normal modes which 
are commonly used to study the dynamics at low temperatures
\cite{ref:Schober723,ref:Laird636,ref:Schober6746,ref:Schober965,ref:CaprionMRS2002,ref:Schober67,ref:Schober11469-Bembenek936-Keyes4651-Buchenau5039}.

The dynamics of glasses out of equilibrium, i.e. systems which have been 
quenched from high
to low temperature, displays additional complexity, since the system
might ``age'' \cite{ref:ageing_review}.
This means that the dynamics depends on the waiting time 
between the quench and the measurement
\cite{ref:Bouchaud243,ref:Struik,ref:Kob4581} which results in 
the violation of the fluctuation dissipation theorem 
\cite{ref:Kob4581,ref:Barrat637}.

Another fruitful approach to gain insight into the dynamics of
supercooled liquids and glasses has been to investigate the energy
landscape of the inherent structure, that means of the instantaneous
configurations which have been quenched to their local potential
energy minimum 
\cite{ref:Doliwa4915,ref:Sastry301,ref:DoliwaS849,ref:Schroeder9834,ref:Doliwa031506-Doliwa0306343,ref:Doliwa030501,ref:SaksaengwijitS1237}.
At the same temperature when the system starts slowing down
drastically, the potential energy of the inherent structure
undergoes a qualitative change \cite{ref:Sastry301}. The long time
behavior of this potential energy shows jumps between metabasins, 
where the latter are groups of strongly correlated local minima
\cite{ref:DoliwaS849,ref:Schroeder9834,ref:Doliwa031506-Doliwa0306343,ref:Doliwa030501,ref:SaksaengwijitS1237}.
The mean average waiting time $\langle \Delta t_\ind{wait} \rangle$
between these jumps turns out to be dominated by the long times.
$\langle \Delta t_\ind{wait} \rangle$ together with the diffusion
constant allow an estimate of the cage size \cite{ref:Doliwa030501}.

Direct studies of the cage have been done via three-time
correlations \cite{ref:Doliwa4915}, velocity-velocity correlations
\cite{ref:Weeks095704,ref:Hurley1694,ref:Weeks361}, via 
the cage correlation function
\cite{ref:Rabani3649,ref:Rabani6867,ref:Gezelter3444}, and 
a passage time before a particle escapes  \cite{ref:Allegrini5714}.
The results of these studies tell us about 
cage properties such as the cage size and waiting time 
within a cage (or jump rate).

Hand in hand with the experiments and simulations are the 
{\bf theoretical models} for supercooled liquids and glasses.
One of the very successful theories is the mode coupling theory 
of the glass transition (MCT)
\cite{ref:Leutheusser2765-Bengtzelius5915},
which describes the dynamics of supercooled liquids via non-linear 
dynamics of coupled density modes and makes predictions for 
quantities such as $D$, $F(q,t)$, and the susceptibility
$\chi''(\omega)$.
The extended version of MCT includes hopping processes via the
coupling to current fluctuations 
\cite{ref:Goetze415-Das2265-Goetze3407-Sjoegren5-Fuchs7709}.
The comparison of experiment and this theory for $\chi''(w)$
\cite{ref:Li43-Goetze4133}  and $\alpha_2$ \cite{ref:Fuchs3384} 
shows very good agreement.

The MCT is a theory for the glass transition for temperatures
above the glass transition. Below the glass transition there exists 
no equivalent of a microscopic theory such as MCT.
Some phenomenological theories which are more general models 
for solids build in hopping either indirectly as 
in the free volume model \cite{ref:Cohen1077} or directly in 
hopping models
\cite{ref:Bendler3970,ref:Chudley353-Dyre7243,ref:Dyre457-Schroeder3173}. 
Direct hopping models are based on the assumption that hopping is
the main process which explains the dynamics.
The starting point in these models is 
usually a random walk and the anomalous
diffusion is incorporated via a distribution of waiting times (or
jump rates). The jumping entities are either defects
\cite{ref:Bendler3970}, uncharged carriers 
\cite{ref:Chudley353-Dyre7243}
or (in the case of predictions for ac and dc conductivity) charged
carriers 
\cite{ref:Dyre457-Schroeder3173}.

The goal of this paper is to obtain via single particle trajectories
direct information about jump processes such as jump sizes and
waiting times between jumps. Specific examples of the
dynamics of single particle jumps, in the form of 
a plot of one component of ${\bf r}_i(t)$ 
\cite{ref:Sanyal361,ref:Wahnstroem3752,ref:Bhattacharyya2741,ref:Miyagawa3879} or in the form of a two- or three-dimensional picture of the 
particle trajectory
\cite{ref:Gaukel1907,ref:Kegel290,ref:Weeks627,ref:Weeks095704,ref:Hurley10521,ref:Doliwa6898,ref:Schroeder9834,ref:Weeks361},
are very helpful to get an idea of some qualitative features of jumps,
such as the detailed geometry of jump processes
\cite{ref:Sanyal361,ref:Wahnstroem3752,ref:Schroeder9834,ref:Miyagawa3879}.
In this paper, however, we go beyond single examples by defining
a systematic search algorithm for jump processes.
For the case of single particle trajectories a similar approach has
been used in the work
\cite{ref:Wahnstroem3752,ref:KlugePhD,ref:Miyagawa8278,ref:Teichler717},
where the jump criterion is a minimum hopping distance,
which is the same for all particles. 
In contrast, we use a relative criterion, where for each
particle its size of fluctuations is compared to its jump size. 
We choose this relative criterion to be able to identify jumps 
of particles of different sizes and neighborhood. That means the 
criterion is adjusted to the cage size of each individual particle.
As criterion for the occurrence of a jump we use the positions 
of particles instead of their energy. While both approaches are 
fruitful, we believe that many theoretical models and 
interpretations of simulations and experiments are based on 
an intuitive picture of the particle motion in real space.
Similarly our goal is to mimic a careful observer of each 
particles motion in our system. We therefore use 
single particle trajectories,
instead of a quantity which is an average over 
all particles (as in the work
\cite{ref:Oligschleger660-811,ref:Teichler339-533,ref:Schober40,ref:Gaukel1907,ref:Caprion3709,ref:Schober723,ref:Sanyal4154,ref:Delaye2063}).
For the case of single particle trajectories the distinction between
vibrational and hopping motion turns out to be clearer
by taking time averages than by using trajectories of the inherent
structure.

In the following we define our model, and give details about the
simulation (Sec.~\ref{sec:model}). Our precise definition of a jump 
is given in
Sec.~\ref{sec:jumpdef}. We find two types of jumps (irreversible and 
reversible). The latter are distinguished for the rest of the paper
(in distinction from 
\cite{ref:Wahnstroem3752,ref:KlugePhD,ref:Miyagawa8278,ref:Teichler717}.)
In Sec.~\ref{sec:nojumps} we count as a function of temperature the
number of jumping particles, and in Sec.~\ref{sec:nostates} the
number of visited different cages. In Sec.~\ref{sec:times}
we investigate the times during a jump and between successive jumps.
The jump size both in position and in potential energy are presented
in Secs.~\ref{sec:jumplengths} and \ref{sec:energy}.
In Sec.~\ref{sec:conclusions} 
we conclude with a summary of our results, a comparison with the
results of previous work and with our resulting picture of jump 
processes. We finish with open questions suggesting future work.
%
\section{Model and Simulation}  
\label{sec:model}

We use a binary Lennard-Jones (LJ) mixture of 800 A and 200
B particles with the same mass. The
interaction potential for particles $i$ and $j$ at 
positions ${\bf r}_i$ and ${\bf r}_j$ and of type 
$\alpha, \beta \in \{$A,B$\}$ is 
\begin{equation}  
\label{eq:Valphabeta}
V_{\alpha \beta}(r) = 4 \, \epsilon_{\alpha \beta} \,
\left ( \left ( \frac{\sigma_{\alpha \beta}}{r} \right )^{12}
      - \left ( \frac{\sigma_{\alpha \beta}}{r} \right )^{6}
\right ),
\end{equation}
where $r = |{\bf r}_i - {\bf r}_j|$ and 
$\epsilon_\ind{AA}=1.0 , \epsilon_\ind{AB}=1.5 , \epsilon_\ind{BB}=0.5 ,
\sigma_\ind{AA}=1.0 , \sigma_\ind{AB}=0.8$ and
$\sigma_\ind{BB}=0.88$. We truncate and shift the potential at 
$r=2.5 \,\sigma_{\alpha \beta}$. 
From previous investigations
\cite{ref:kob_andersenI,ref:kob_andersenII} it is known
that this system is not prone to crystallization and demixing. 
In the following we
will use reduced units where the unit of length is $\sigma_\ind{AA}$,
the unit of energy is $\epsilon_\ind{AA}$ and the unit of time is 
$\sqrt{m \sigma_\ind{AA}^2/(48 \epsilon_\ind{AA})}$. 

We carry out molecular dynamics (MD) simulations using the velocity
Verlet algorithm with a time step of 0.02. The volume is kept constant
at $V=9.4^3=831$ and we use periodic boundary conditions. 
According to previous simulations
\cite{ref:Barrat637,ref:kob_andersenI,ref:kob_andersenII} the 
kinetic glass transition is around $T \approx 0.435$. We analyze here 
simulations at $T=0.15/0.2/0.25/0.30/0.35/0.38/0.40/0.41/0.42$ and 
$0.43$ as they have been described in
\cite{ref:kvldynhetJCP}. We start with 10 independent well
equilibrated configurations at $T=0.446$.\cite{ref:initconf}
Each of these
configurations undergoes the following sequence of simulation
runs. After an instantaneous quench to $T=0.15$ we first run a (NVT)
simulation for $10^5$ MD steps. The temperature is kept constant by
replacing the velocities of all particles by new
velocities drawn from the corresponding Boltzmann distribution 
every 50 time steps. We
then run the simulation for $10^5$ MD steps without the temperature
bath (NVE) and subsequently raise the temperature instantaneously to
the next higher temperature $T=0.2$. (NVT) and (NVE) runs of 
$10^5$ MD steps each follow and the temperature is again raised,
and so forth. The final configurations
of these (NVT) runs are the initial configurations for the (NVE)
production runs (for $5 \cdot 10^6$ MD steps) presented in this
paper. During each production run the positions of all particles
(configurations) are stored every 2000 MD steps which are then
used for analysis.

For the here investigated temperatures the relaxation times $\tau$ are 
much larger than the waiting time before the production runs 
(at $T = 0.446$ is $\tau \approx 8 \cdot 10^5$). We therefore study 
relaxation processes out of equilibrium and find aging effects, 
which is consistent with previous detailed studies of aging 
of the same binary Lennard-Jones system
\cite{ref:Kob4581,ref:Barrat637}.

\section{Definition of Jump and Jump-Type} \label{sec:jumpdef}

We focus in this paper on the process of a particle escaping its
cage, using single particle 
trajectories ${\bf r}_i(t)$ given by the periodically 
stored configurations.

\begin{figure}[htb]
\epsfxsize=3.2in
\epsfbox{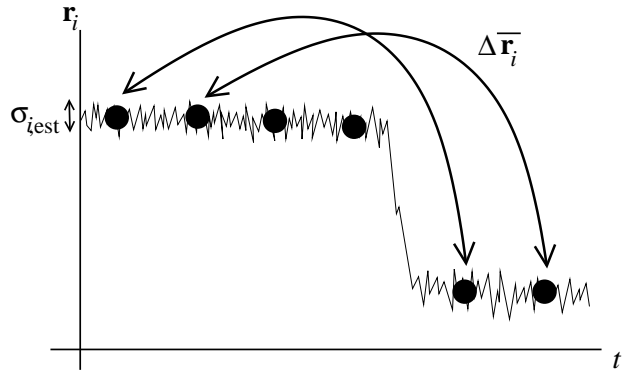}
\caption{ Sketch of a particle trajectory to illustrate the 
  definition of a jump by comparison of the fluctuations of 
  the particle $\sigma_{i,\ind{est}}$ and the difference in 
  average positions $\Delta \overline{{\bf r}_i}$.
}
\label{fig:jumpdef}
\end{figure}

To distinguish vibrations around an average position from a 
change in the average position, we average 20 consecutive
positions ${\bf r}_i(t)$ to obtain 
$\overline{{\bf r}_i}(m)$ as sketched in Fig.~\ref{fig:jumpdef}.
We identify jumps by comparing changes in these averaged positions 
$|\Delta \overline{{\bf r}_i}|$ with the fluctuations in 
position $\sigma_{i,\ind{est}}$ for each particle $i$ where
\begin{equation}
\label{eq:deltaoverlri}
|\Delta \overline{{\bf r}_i}| = |\overline{{\bf r}_i}(m) - 
   \overline{{\bf r}_i}(m - 4*20)|
\end{equation}
(see Fig.~\ref{fig:jumpdef}) and $\sigma_{i,\ind{est}}$ is defined 
in \cite{ref:refnoteIII1}.
We use in specific for $|\Delta \overline{{\bf 
r}_i}|$ not consecutive but instead average positions 
which are four averages (each of 20 configurations) apart.
This choice of time interval in $|\Delta \overline{{\bf r}_i}|$
is to allow identification of not
only sudden but also more gradual jumps.
We define that a jump of particle $i$ occurs whenever
\begin{equation}
\label{eq:jumpoccur}
|\Delta \overline{{\bf r}_i}|^2 > 20\sigma^2_{i,\ind{est}}
\end{equation}
Notice that we use a relative criterion, namely for each particle $i$ 
a comparison of $|\Delta \overline{{\bf r}_i}|$ with its 
$\sigma_{i,\ind{est}}$. 
Our motivation for this relative criterion is that we would like to 
identify jumps of both A and the smaller B particles. Also
even for particles of the same type their jump size might differ 
due to different cage sizes. $\sigma^2_{i,\ind{est}}$ is an 
estimate for the cage size of each individual particle and 
is therefore used as criterion for the occurrence of a jump.
\clearpage

\begin{figure}[htb]
\epsfxsize=3.2in
\epsfbox{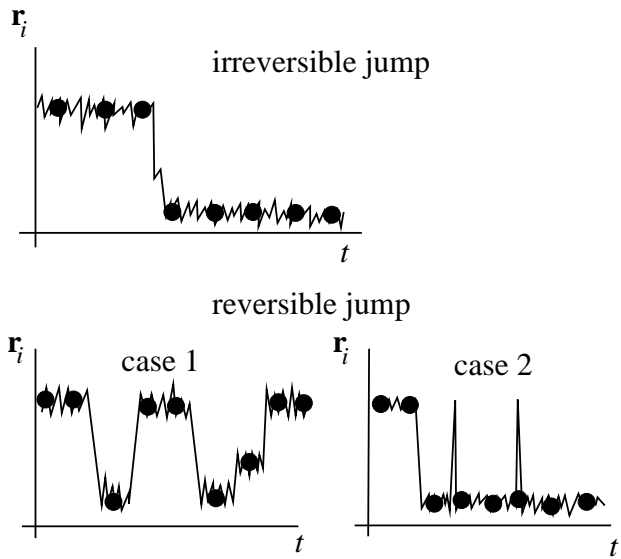}
\caption{ Sketch of typical particle trajectories to illustrate the 
  distinction between irreversible and reversible jumps
}
\label{fig:jumptype}
\end{figure}

When we apply the above jump definition, we find two types of 
jumps which we call ``irreversible'' and ``reversible'' jumps 
(see Fig.~\ref{fig:jumptype}).  
A particle which 
undergoes an irreversible jump succeeds in escaping its cage (for the time
window of the simulation) whereas a particle undergoing a reversible jump 
returns back to one of its previous average positions. Similar results
have been found in previous simulations of other systems
\cite{ref:Oligschleger660-811,ref:Teichler339-533,ref:Gaukel1907,ref:Schober723,ref:Wahnstroem3752,ref:Gaukel664,ref:Miyagawa3879}.
However, the present work differs from these that
we analyze all following
quantities for the reversible and irreversible jumps separately. As sketched
in Fig.~\ref{fig:jumptype}, we call a jump reversible whenever case~1
or case~2 occurs. Case~1 corresponds to the situation of a
particle undergoing multiple jumps and returning to one or more 
of any previous average positions 
(for details see Sec.~\ref{sec:nostates}). 
If two average positions are equal then all jumps 
which happen between the previous position and the recurring position 
are called reversible jumps. In the example of
Fig.~\ref{fig:jumptype} this means in case~1 that all shown 
jumps are reversible jumps. To increase our resolution in time we
use not only the information of the averaged positions
$ \overline{{\bf r}_i}(m)$ but also the complete information of 
${\bf r}_i(t)$. In case~2 (see Fig.~\ref{fig:jumptype})
the spikes in ${\bf r}_i(t)$ indicate 
returns to the average position before the jump \cite{ref:refnoteIII3}.
If a jump satisfies neither case~1 nor case~2, then it is called
``irreversible jump.'' 

In the following we distinguish between the 
jump events, as they have been defined so far,
and the corresponding jumping particles which 
may undergo multiple jumps of different types. Any 
reversible jump designates the corresponding particle as 
a reversible jumper for the entire time window.

\section{Number of Jumping Particles} \label{sec:nojumps}

\begin{figure}[htb]
\epsfxsize=3.2in
\epsfbox{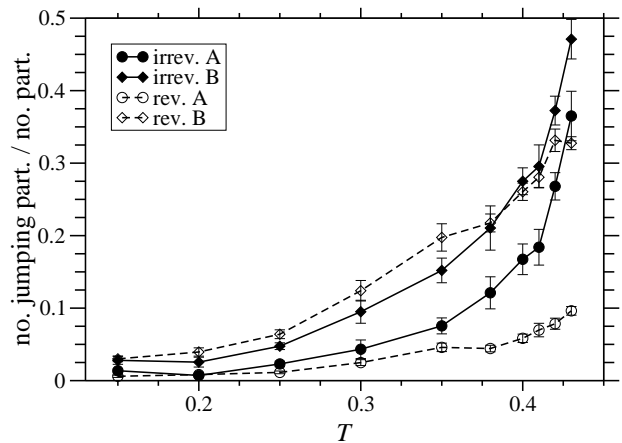}
\caption{ Number of jumping particles normalized by the number 
of corresponding particle type as a function of temperature $T$. 
Irreversible and reversible jumpers are distinguished.
}
\label{fig:nojumpsnorm}
\end{figure}

We apply now the above definitions 
to identify all jumps occurring in our simulations. 
The number of identified jumps depends on the time interval of the 
production runs and, since the system is out of equilibrium, also on the 
waiting time before the production runs.
The straightforward counting of jump events would give
reversible jumpers (which sometimes jump many times)
a larger weight. 
We therefore present 
the number of jumping particles. Fig.~\ref{fig:nojumpsnorm} shows 
the number of jumping particles normalized by the number $N_{\alpha}$ 
of particles in the system of type $\alpha \in \{\mbox{A,B}\}$.
With increasing temperature $T$ the number 
of jumping particles increases consistent with an increasing number 
of relaxation processes. 
A similar temperature dependence has been found indirectly via the
participation ratio in the work 
\cite{ref:Oligschleger660-811,ref:Schober67,ref:Oligschleger1031}.
More surprisingly, we find that not only the
small B-particles are jumping but also a considerable fraction of A
particles. However, the fraction of jumping B
particles is larger than the fraction of jumping A particles due to the
smaller size and therefore higher mobility of the B particles.
Furthermore, both irreversible and reversible jumps are not only occurring in
a certain temperature range but at all temperatures. 

\clearpage

\begin{figure}[htb]
\epsfxsize=3.2in
\epsfbox{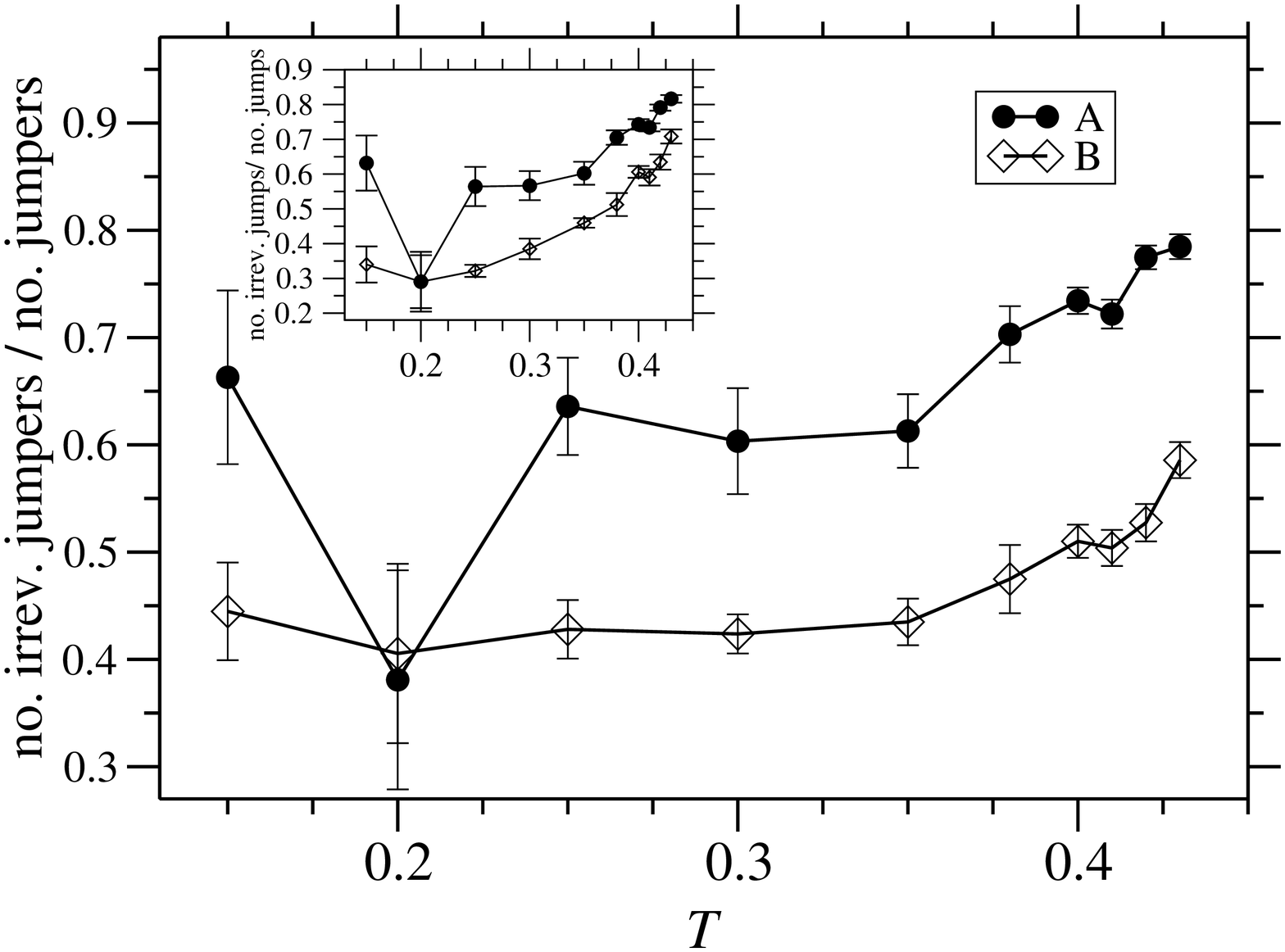}
\caption{ Number of irreversible jumping particles divided by the number
of both reversible and irreversible jumping particles. 
The inset shows the number of irreversible 
jump {\em events} divided by the number of jump events.
}
\label{fig:norealdivboth}
\end{figure}
\begin{figure}[htb]
\epsfxsize=3.2in
\epsfbox{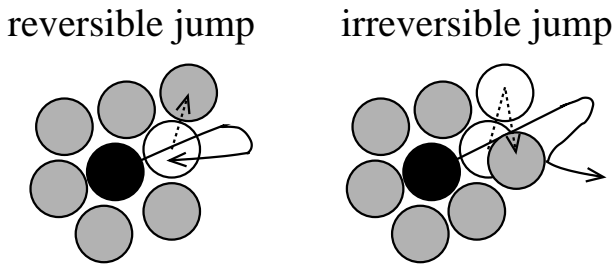}
\caption{ This picture illustrates our interpretation of 
irreversible and reversible jumps and Fig.~\ref{fig:norealdivboth}.
}
\label{fig:met_real_picture}
\end{figure}

Fig.~\ref{fig:norealdivboth} illustrates the fraction of these jumping
particles that are irreversible jumpers. 
At low to intermediate temperatures this
ratio is, within the large error bars, roughly constant. 
At intermediate to larger temperatures 
irreversible jumpers become more likely than reversible jumpers
with increasing temperature.
This increase is even more pronounced for all temperatures 
if one considers the number of irreversible jump {\em events} (instead 
of jumping particles) divided by 
the number of all jump events (see inset of
Fig.~\ref{fig:norealdivboth}). 
Gaukel {\it et al.}\ \cite{ref:Gaukel1907,ref:Schober723,ref:Gaukel664}
come to a similar conclusion via a model for their simulation data.
Their irreversible jumps become more likely with increasing 
temperature.
We interpret this increase in the
fraction of irreversible jumps as sketched in
Fig.~\ref{fig:met_real_picture}. 
Both irreversible and reversible jumps
start out the same, with the jumping particle leaving
the cage (formed by the neighboring particles) possibly through an 
opening in the cage. In the case of the irreversible jump, the
entrance of the cage gets meanwhile blocked
by a particle, loosely speaking the door gets closed, and the jumping
particle can no longer return and has successfully escaped its cage.
With increasing temperature all particles become more mobile, which
increases the likelihood of the blockage of the entrance back into
the cage which in turn leads to an increase in the fraction of irreversible
jumps as shown in Fig.~\ref{fig:norealdivboth}.
We do not intend to make here a statement about the exact geometric 
process, for example that a door made up of a single particle gets
open and closed, but more generally the process of
rearrangement of the cage.
Interestingly enough, this blockage happens more
often in the case of the larger A than the smaller B-particles 
which leads to larger ratios for A than B particles in 
Fig.~\ref{fig:norealdivboth}. 

\section{Number of Average Positions} \label{sec:nostates}

With each jump a particle either returns to one of its former
average positions within a cage or to a new overall average
position. 
We call the average positions  before 
and after a jump $\langle \overline{{\bf r}_i} \rangle_\ind{i}$ and 
$\langle \overline{{\bf r}_i} \rangle_\ind{f}$
(for details about the time averages $\langle \cdot \rangle_\ind{i,f}$
see Fig.~\ref{fig:dRifdRmaxdRba_def} and Sec.~\ref{sec:jumplengths}).
We use as criterion for two average positions to be the same that the 
distance between them 
$\Delta \langle \overline{ r_i} \rangle =\left | 
     \langle \overline{{\bf r}_i} \rangle_\ind{f} - 
    \langle \overline{{\bf r}_i} \rangle_\ind{i} \right |$ 
and the average fluctuations before the jump 
$\langle \sigma_i^2 \rangle_\ind{i}$ (for the definition of 
$\sigma_i^2 $ see \cite{ref:refnoteIII1}) satisfy 
$\left ( \Delta \langle \overline{ r_i} \rangle \right )^2 
      \le 5 \langle \sigma_i^2 \rangle_\ind{i}$.
\begin{figure}[htb]
\epsfxsize=3.2in
\epsfbox{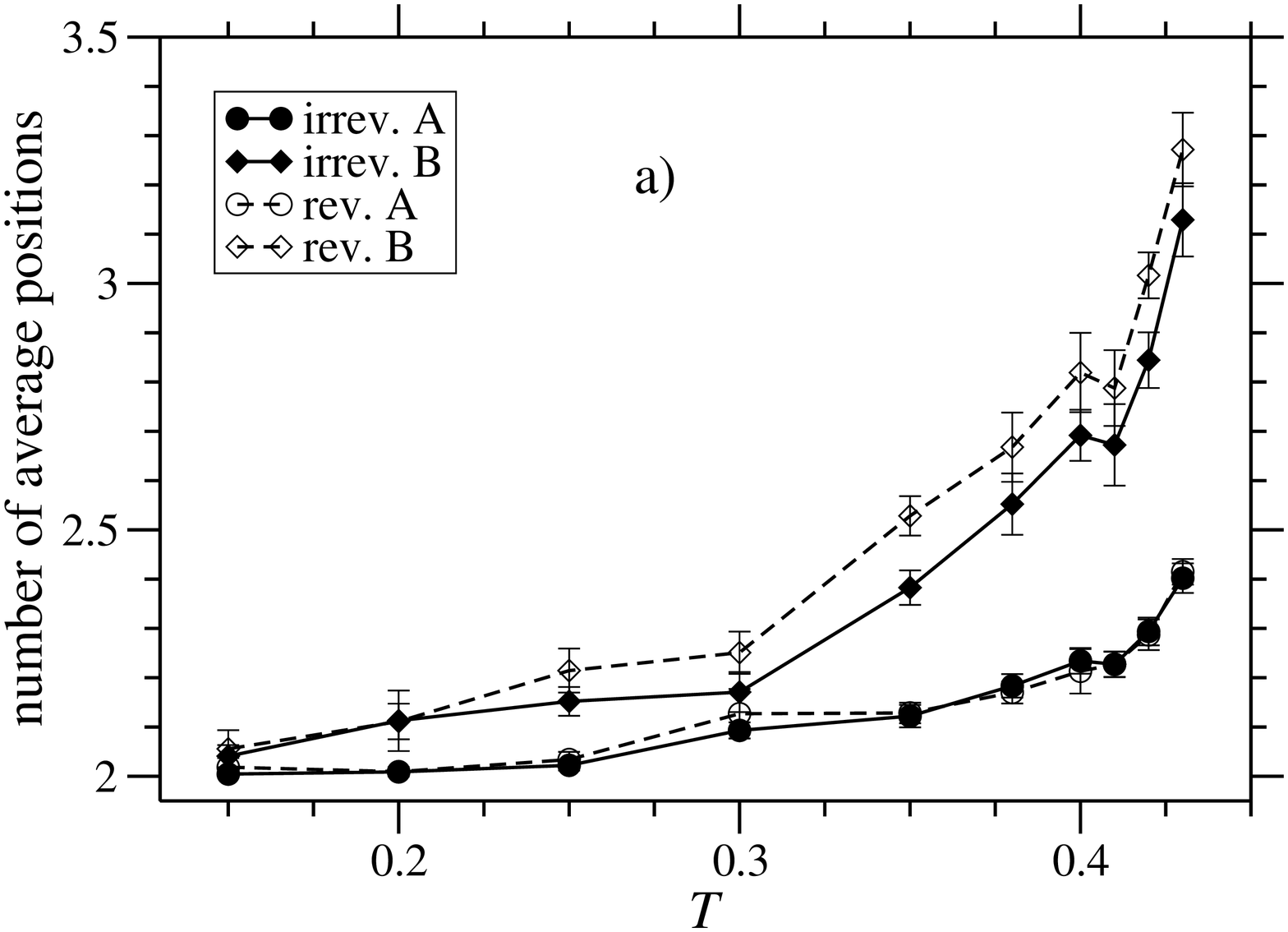}
\epsfxsize=3.2in
\epsfbox{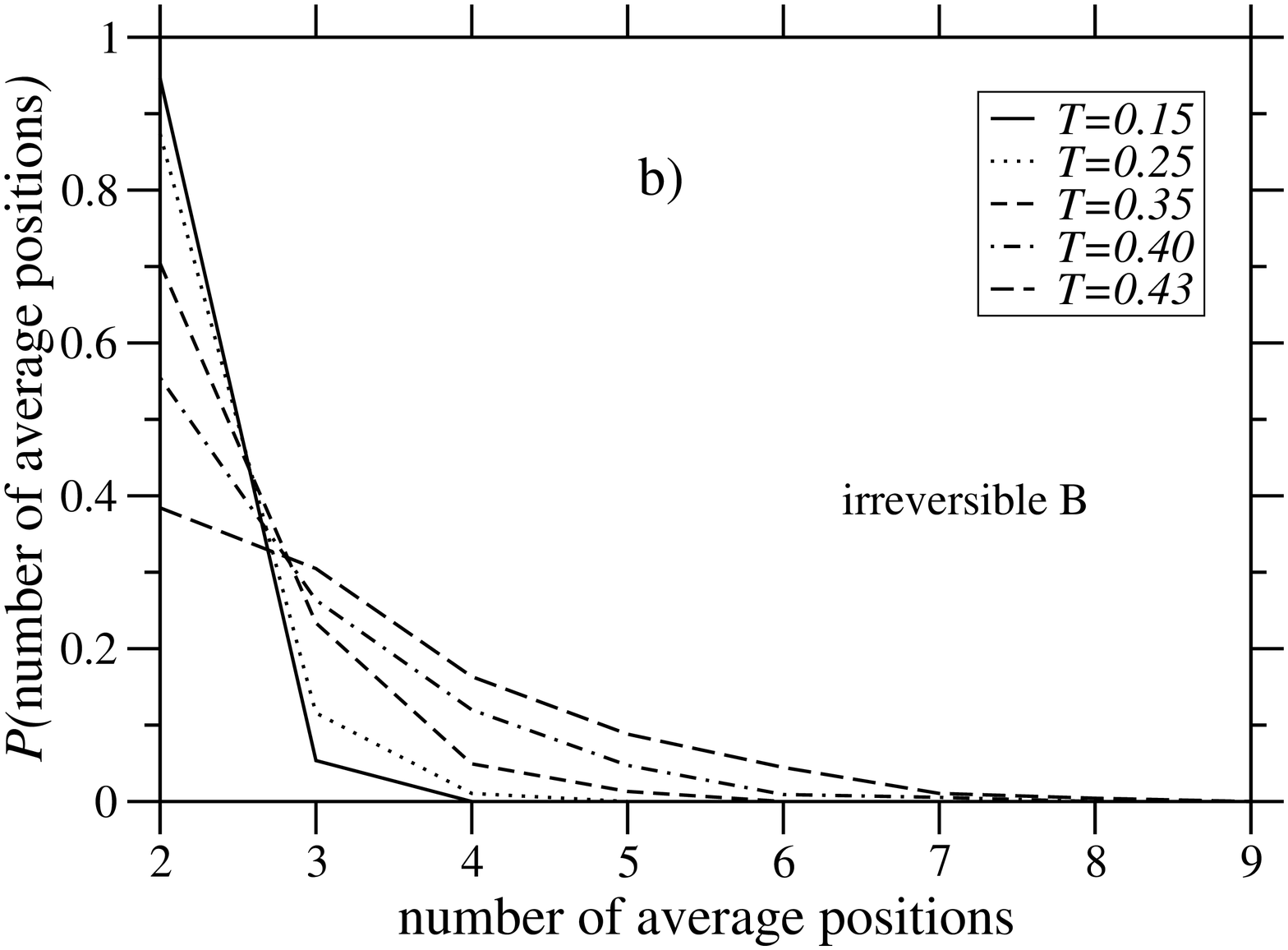}
\caption{ The number of average positions 
(as defined in Sec.~\ref{sec:nostates})
visited by jumping particles in a) as a function of temperature
and in b) its distribution for irreversible jumping B particles.
}
\label{fig:nostates_Pofnostates}
\end{figure}
Fig.~\ref{fig:nostates_Pofnostates}a shows
an average of the number of distinct average positions which are visited
by a particle. According to the higher mobility of the smaller B
particles, they visit more average positions than the A particles. For low
temperatures most jumping particles visit only 
two average positions during our
simulation. For intermediate to larger temperatures, however, not
only increasingly more particles jump (see Sec.~\ref{sec:nojumps}) but
these particles also jump more often. The distribution of the number
of different visited average positions 
at the highest temperatures (as shown in
Fig.~\ref{fig:nostates_Pofnostates}b), broadens with increasing $T$ such
that some particles visit up to seven different average positions
during the simulation run.

\section{Times} \label{sec:times}

\begin{figure}[htb]
\epsfxsize=3.2in
\epsfbox{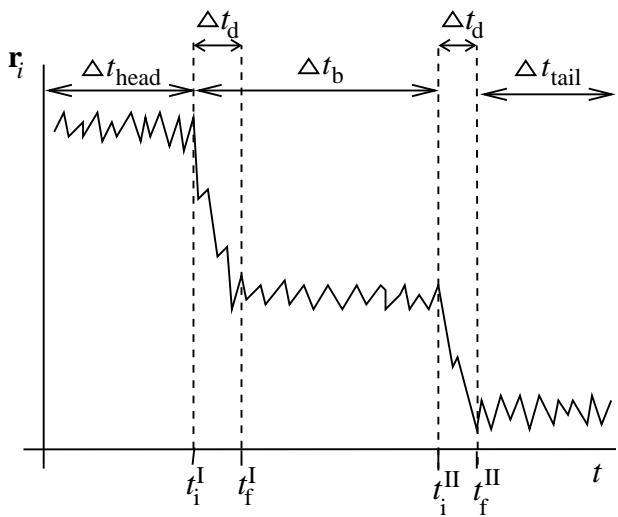}
\caption{ Sketch to illustrate the definitions of the 
starting time $t_\ind{i}$ and ending time $t_\ind{f}$ of a jump
(here for jumps I and II) and the times before the first 
jump of a particle $\Delta t_\ind{head}$, during a jump
$\Delta t_\ind{d}$, between successive jumps $\Delta t_\ind{b}$ and 
after the last jump $\Delta t_\ind{tail}$.
}
\label{fig:tbd_def}
\end{figure}
In this section we investigate the time scale of jumps. As sketched
in Fig.~\ref{fig:tbd_def} we determine the time duration of a
jump $\Delta t_\ind{d} = t_\ind{f} - t_\ind{i}$, the time 
before the first jump of a particle $\Delta t_\ind{head}$, the 
time after the last jump of a particle $\Delta t_\ind{tail}$, and the time 
between two successive jumps I and II to be 
$\Delta t_\ind{b}=t_\ind{i}^{II} -t_\ind{i}^{I}$ where $t_\ind{i}$ 
and $t_\ind{f}$ indicate the starting and ending time of a 
jump \cite{ref:refnoteVI1}. Notice that $\Delta t_\ind{b}$ is only
defined if a jump particle jumps twice or more and that we analyze
$\Delta t_\ind{head}$ and $\Delta t_\ind{tail}$ separately as presented
shortly.  For the distinction of irreversible and reversible jumps, we assign  
the jump type of $\Delta t_\ind{b}$ according to the jump ending
$\Delta t_\ind{b}$, for example in  Fig.~\ref{fig:tbd_def} 
the jump type of $\Delta t_\ind{b}$ is determined by jump II.
This means that $\Delta t_\ind{b}$ is a measure of the waiting time 
{\em before} a jump occurs. 
\begin{figure}[htb]
\vspace*{3mm}
\epsfxsize=3.2in
\epsfbox{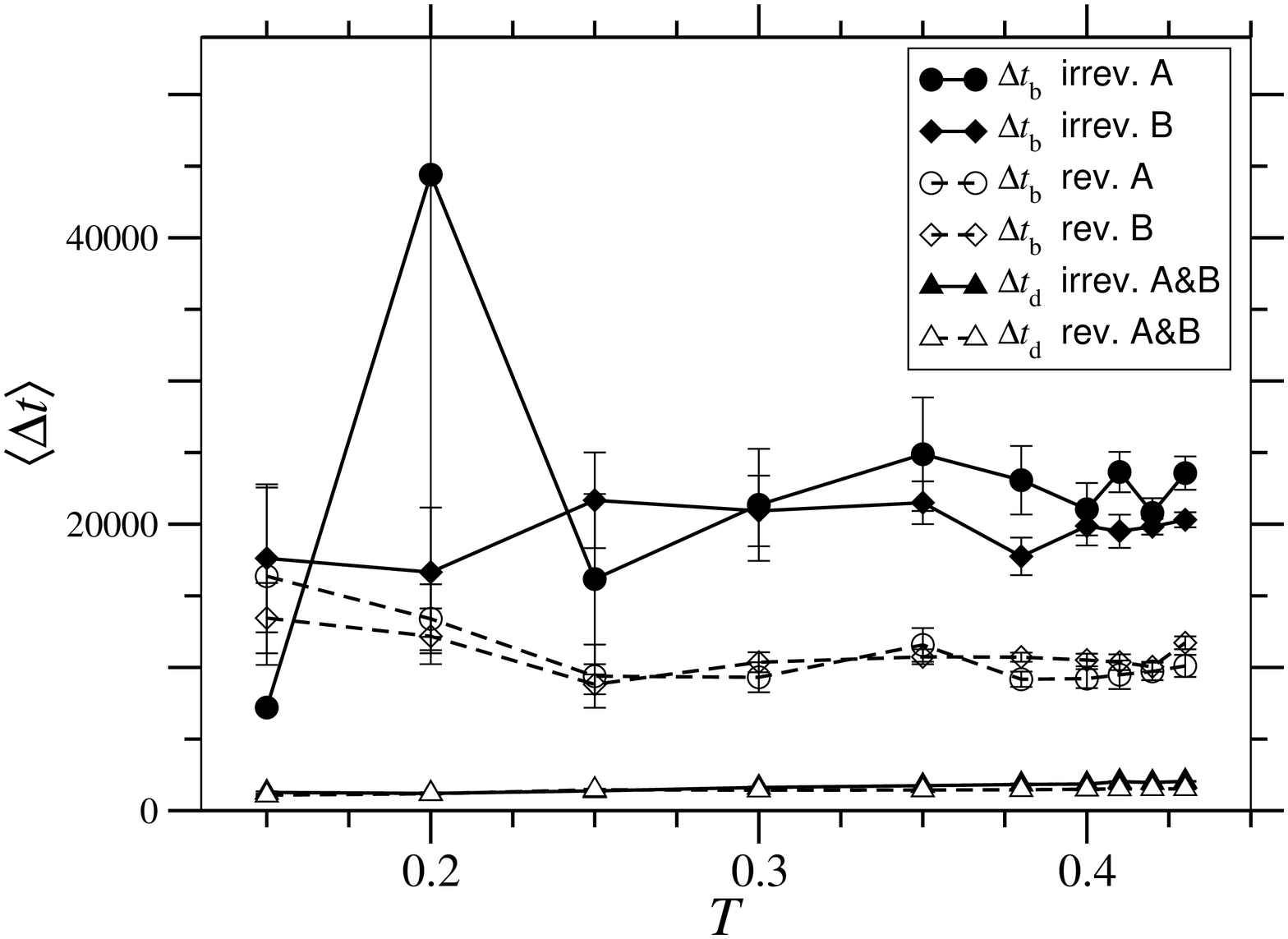}
\vspace*{3mm}
\caption{ Times between jumps $\Delta t_\ind{b}$ for A (circles)
and B (diamonds) particles  and times during jumps
$\Delta t_\ind{d}$ for both A and B particles. 
}
\label{fig:times_sim}
\end{figure}
Fig.~\ref{fig:times_sim} illustrates that 
$\Delta t_\ind{d} \ll \Delta t_\ind{b}$ which tells us in 
hindsight why we could identify jumps as rare events with the above described 
procedure. Since the time resolution is of the order of 1000
($\approx 20 \cdot 2000 \cdot 0.02$), i.e. of the same order 
as $\Delta t_\ind{d}$, we do not draw further conclusions about 
the temperature dependence of $\Delta t_\ind{d}$. 
However, $\Delta t_\ind{b}$ 
is well above our time resolution and small enough to be
detected during our simulation run of length $1 \cdot 10^5$. In
accordance with the picture of reversible jumpers which try but do
not succeed to escape their cage, these trials happen on a shorter
time scale than irreversible jumps. However, to our surprise, 
$\Delta t_\ind{b}$ seems to be independent not only of the particle
type but also of temperature. 
The question arises if this temperature independence of $\Delta t_\ind{b}$ 
is a consequence of the finite time window of our simulation. To take
this finite time interval $t_\ind{tot}$ into account, we make a crude
approximation to correct the time intervals $\Delta t_\ind{b}$,
$\Delta t_\ind{head}$ and $\Delta t_\ind{tail}$ by
assuming that jumps happen
equally likely at any time in the window $[0,t_\ind{tot}]$ (which
is not accurate due to aging). The probability 
$P_\ind{corr}(\Delta t)$ of finding the complete interval $\Delta t$ 
reduces to the probability of finding 
$\Delta t$ during $[0,t_\ind{tot}]$ in the simulation
\begin{equation}
P_\ind{sim}(\Delta t) = P_\ind{corr}(\Delta t) \, \cdot 
           \frac{(t_\ind{tot} - \Delta t)}{t_\ind{tot}} \, \cdot c
\end{equation}
where $c$ is a normalization constant. We may approximate 
$c$ with ${\left \langle \frac{t_\ind{tot}}{t_\ind{tot} - \Delta t}
\right \rangle }_\ind{sim}^{-1}$ where $\langle \cdot \rangle_\ind{sim}$ indicates
an average over jump events of the simulation. We thus obtain 
\begin{equation}
\label{eq:Pcorr}
P_\ind{corr}(\Delta t) \approx \frac{P_\ind{sim}(\Delta t)}
                                    {(t_\ind{tot} - \Delta t) }
   \cdot \left \langle 
        \frac{1}{(t_\ind{tot} - \Delta t)} \right \rangle_\ind{sim}^{-1}
\end{equation}
Similarly the average times $\langle \Delta t \rangle_\ind{sim}$ may be 
approximately corrected by 
\begin{equation}
\label{eq:avdtcorr}
\langle \Delta t \rangle_\ind{corr} \approx \left \langle \frac{\Delta t}
            {(t_\ind{tot}-\Delta t)} \right \rangle_\ind{sim} \cdot
\left \langle \frac{1}{(t_\ind{tot}-\Delta t)}\right \rangle_\ind{sim}^{-1}
\end{equation}
\clearpage
%
\begin{figure}[htb]
\epsfxsize=3.2in
\epsfbox{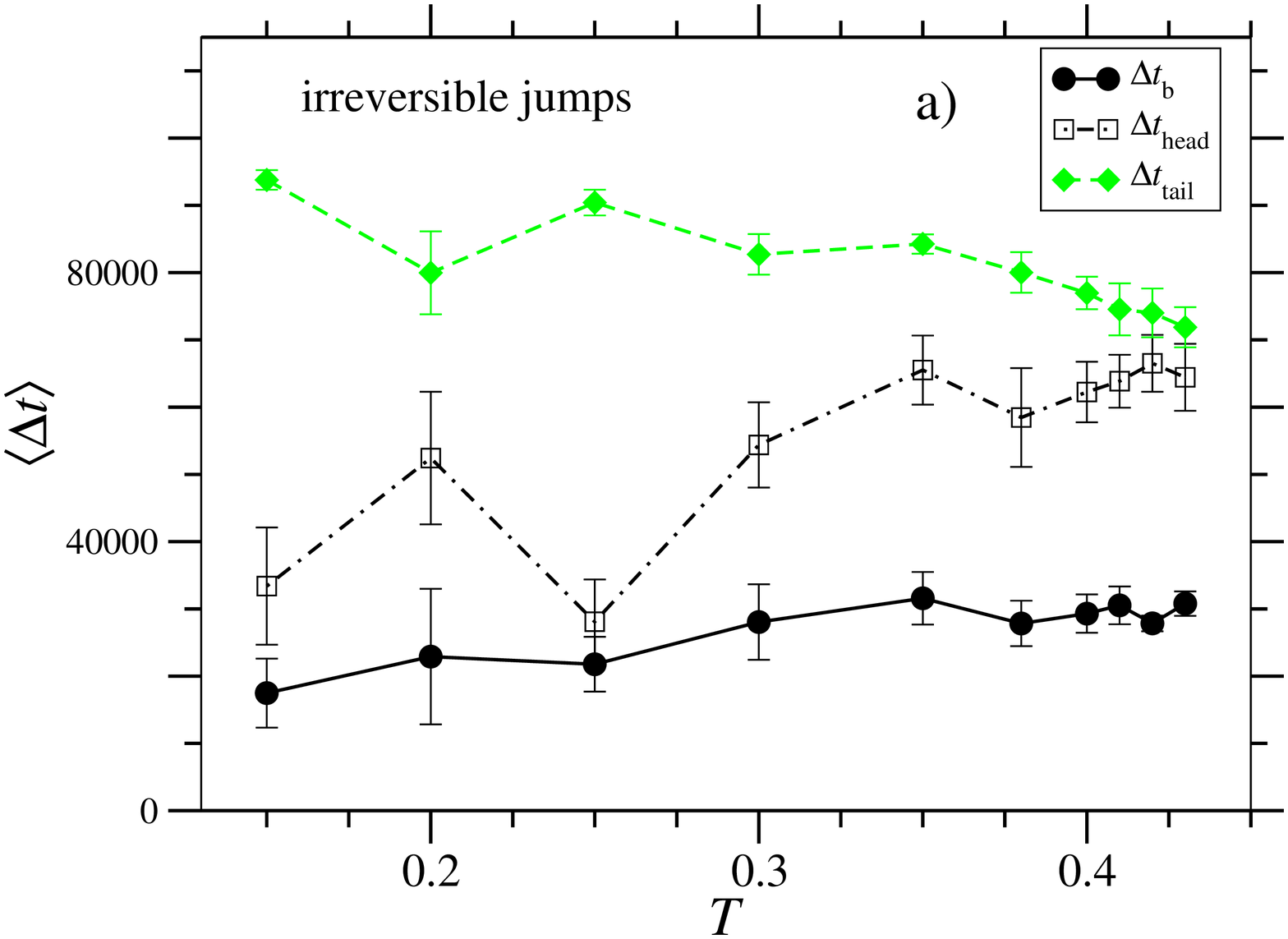}
\epsfxsize=3.2in
\epsfbox{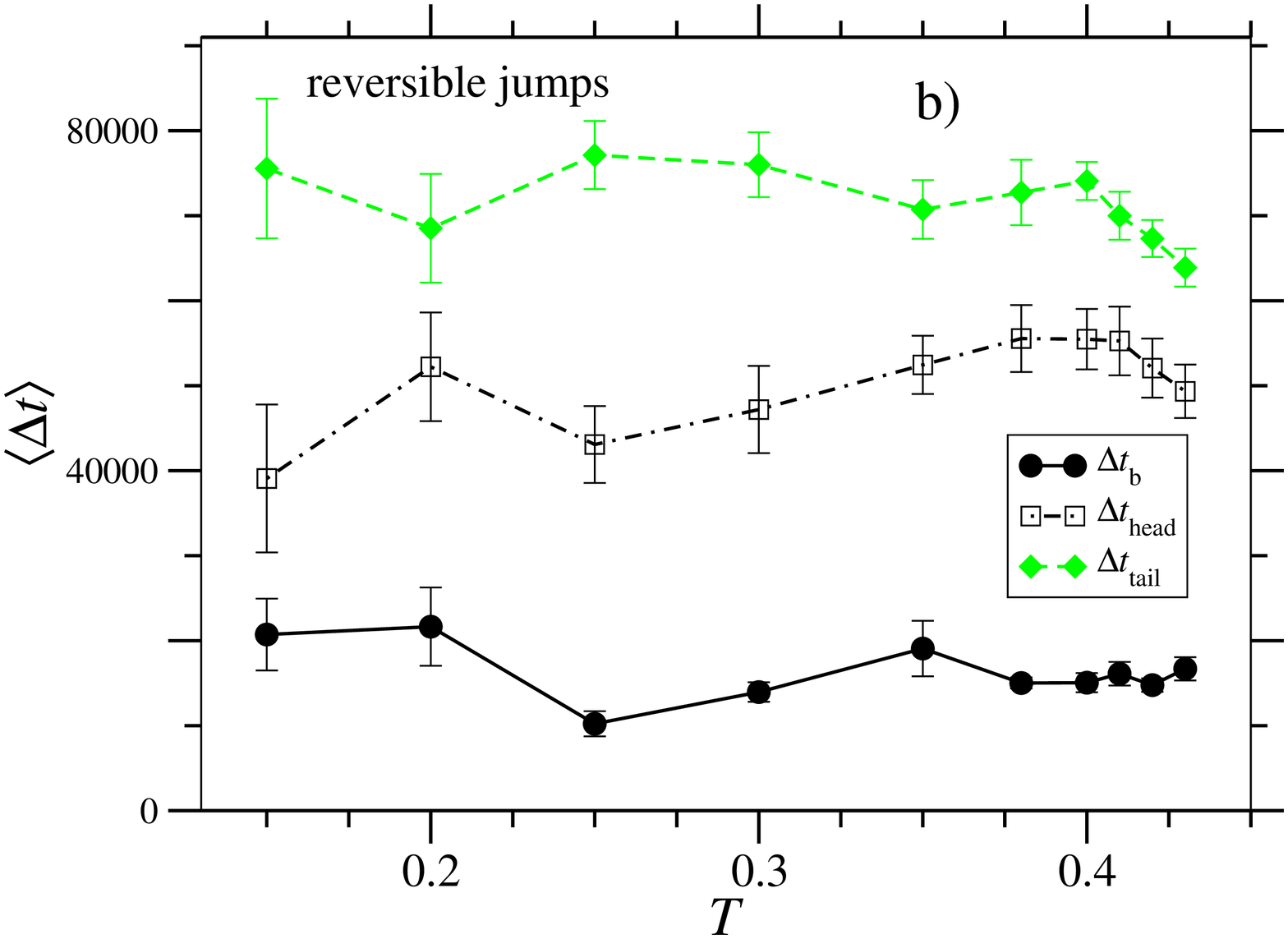}
\vspace*{3mm}
\caption{ Times  $\Delta t_\ind{b}$, $\Delta t_\ind{head}$ and 
$\Delta t_\ind{tail}$ a) for irreversible jumps and b) for 
reversible jumps of A and B particles
using Eq.~(\ref{eq:avdtcorr}).
}
\label{fig:times_corr_real_met}
\end{figure}
Fig.~\ref{fig:times_corr_real_met}a and 
Fig.~\ref{fig:times_corr_real_met}b show the 
resulting corrected times $\langle \Delta t \rangle_\ind{corr}$ 
for irreversible and reversible jumps respectively. For 
simplification of notation we drop $\langle \cdot \rangle_\ind{corr}$
in the following. The tail and head times, $\Delta
t_\ind{tail}$ and $\Delta t_\ind{head}$, reveal that we find aging,
since $\Delta t_\ind{tail} > \Delta t_\ind{head}$, i.e. jumps are
more likely to occur at the beginning of the simulation than later.
This particular aging effect decreases with increasing temperature.
$\Delta t_\ind{tail}$ and $\Delta t_\ind{head}$ are both larger than 
$\Delta t_\ind{b}$ because $\Delta t_\ind{b}$ includes only times of
particles which jump multiple times whereas $\Delta t_\ind{head}$
and $\Delta t_\ind{tail}$ include also times of particles which jump
only once. 
Interestingly, $\Delta t_\ind{b}$ is (although approximately
corrected) still independent of temperature. 
Together with our results of Secs.~\ref{sec:jumpdef} --
\ref{sec:nostates} we therefore find that with increasing
temperature relaxation processes are accelerated via
more jumping particles and more multiple
jumps, but the times between multiple jumps do not become shorter.
This is contrary to previous results 
\cite{ref:KlugePhD,ref:Miyagawa8278,ref:Doliwa031506-Doliwa0306343,ref:Doliwa030501,ref:Vardeman2568} 
in which waiting times decrease with increasing temperature.
In our system, however, the temperature 
independence seems to be true not only for the
averages but even for the distribution $P_\ind{corr}(\Delta t_\ind{b})$
(see Fig.~\ref{fig:Poftb_v2}).
\begin{figure}[htb]
\epsfxsize=3.2in
\epsfbox{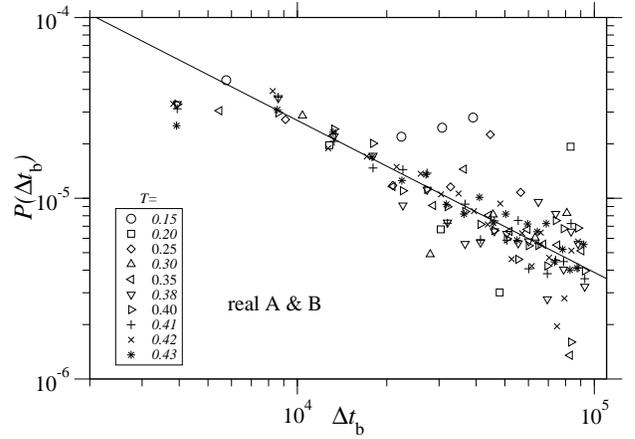}
\vspace*{3mm}
\caption{ Log-log plot of $P(\Delta t_\ind{b})$
using Eq.~(\ref{eq:Pcorr}) for irreversible jumps 
of A and B particles. The line is a linear fit to the data 
with slope $-0.84$.
}
\label{fig:Poftb_v2}
\end{figure}
We interpret this temperature independence as being due to aging which
is consistent with our data for $\Delta t_\ind{tail}$ and 
$\Delta t_\ind{head}$ and with the results of Doliwa {\it et al.} 
\cite{ref:Doliwa031506-Doliwa0306343,ref:Doliwa030501} who
find that the distribution of waiting times initially is temperature 
independent and becomes temperature dependent at later times.
As shown in Fig.~\ref{fig:Poftb_v2} for irreversible jumps of A and B 
particles, $P(\Delta t_\ind{b})$ 
approximately follows a power law $P(\Delta t_\ind{b}) \propto \Delta
t_\ind{b}^{- \nu}$ with $\nu = 0.84$. Even though we expect the system
to show normal diffusion on a time scale much larger than our
simulation run, the intermediate dynamics is subdiffusive 
\cite{ref:eweeks_subdiff}. Please notice however that a direct
comparison of our result with the scheme presented in 
Fig.~1 of \cite{ref:eweeks_subdiff} 
is not possible because in our case the multiple jumps of one particle
at different times as well as the jumps of different particles are not
independent, which changes the dynamics \cite{ref:fogleman}.

\begin{figure}[htb]
\epsfxsize=3.2in
\epsfbox{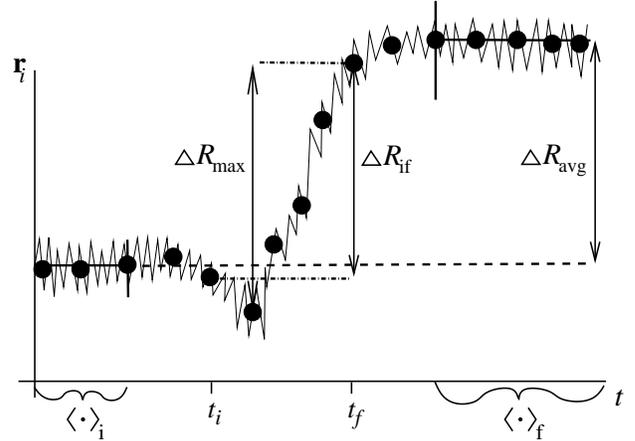}
\vspace*{1mm}
\caption{ Sketch for the definitions of $\Delta R_\ind{if}$,
$\Delta R_\ind{max}$ and $\Delta R_\ind{avg}$.
}
\label{fig:dRifdRmaxdRba_def}
\end{figure}
\clearpage
\section{Jump Lengths} \label{sec:jumplengths}

In the last sections we have investigated how many particles jump 
and how often jumps occur. Next we study how far these particles jump.
To be able to test if our qualitative results are dependent on the 
definition of jump length, we use three quantities: $\Delta R_\ind{if}$, 
$\Delta R_\ind{max}$ and $\Delta R_\ind{avg}$ as sketched in 
Fig.~\ref{fig:dRifdRmaxdRba_def}. 
For a jump of 
particle $i$ the jump distance of the
jump starting at $t_\ind{i}$ and ending at $t_\ind{f}$ (for the 
definition of $t_\ind{i}$ and $t_\ind{f}$ see \cite{ref:refnoteVI1}) is
\begin{equation}
\Delta R_\ind{if} = \left | \overline{\bf{r}_i}(t_\ind{i}) 
                  - \overline{\bf{r}_i}(t_\ind{f}) \right |
\end{equation}
and the maximal distance being detected during the jump
\begin{equation}
\Delta R_\ind{max} = \max_t \left | \overline{\bf{r}_i}(t) 
                  - \overline{\bf{r}_i}(t-3200) \right |
\end{equation}
(here $3200 = 4 \cdot 20 \cdot 2000 \cdot 0.02$ as 
in \cite{ref:refnoteIII1} and Sec.~\ref{sec:jumpdef}) where $t$ is varied 
over times $t$ which satisfy
$t_\ind{i,detect} \le t < t_\ind{f,detect}$ 
(\cite{ref:refnoteVI1}).
The third length $\Delta R_\ind{avg}$ is less dependent on fluctuations 
than $\Delta R_\ind{if}$ and $\Delta R_\ind{max}$.
It is the distance of the overall average positions before and 
after the jump
\begin{equation}
\Delta R_\ind{avg} = \left | \left \langle \overline{\bf{r}_i} 
                            \right \rangle_\ind{f}
              -  \left \langle \overline{\bf{r}_i} 
                            \right \rangle_\ind{i} \right |
\end{equation}
For the averages $\langle \cdot \rangle_\ind{f}$ and 
$\langle \cdot \rangle_\ind{i}$ the positions 
$\overline{{\bf r}_i}(t)$ are averaged over times before and after the
jump excluding a broadened time window around the jump 
$[t_\ind{i} - 800, t_\ind{f}+800]$. 
\begin{figure}[htb]
\epsfxsize=3.2in
\epsfbox{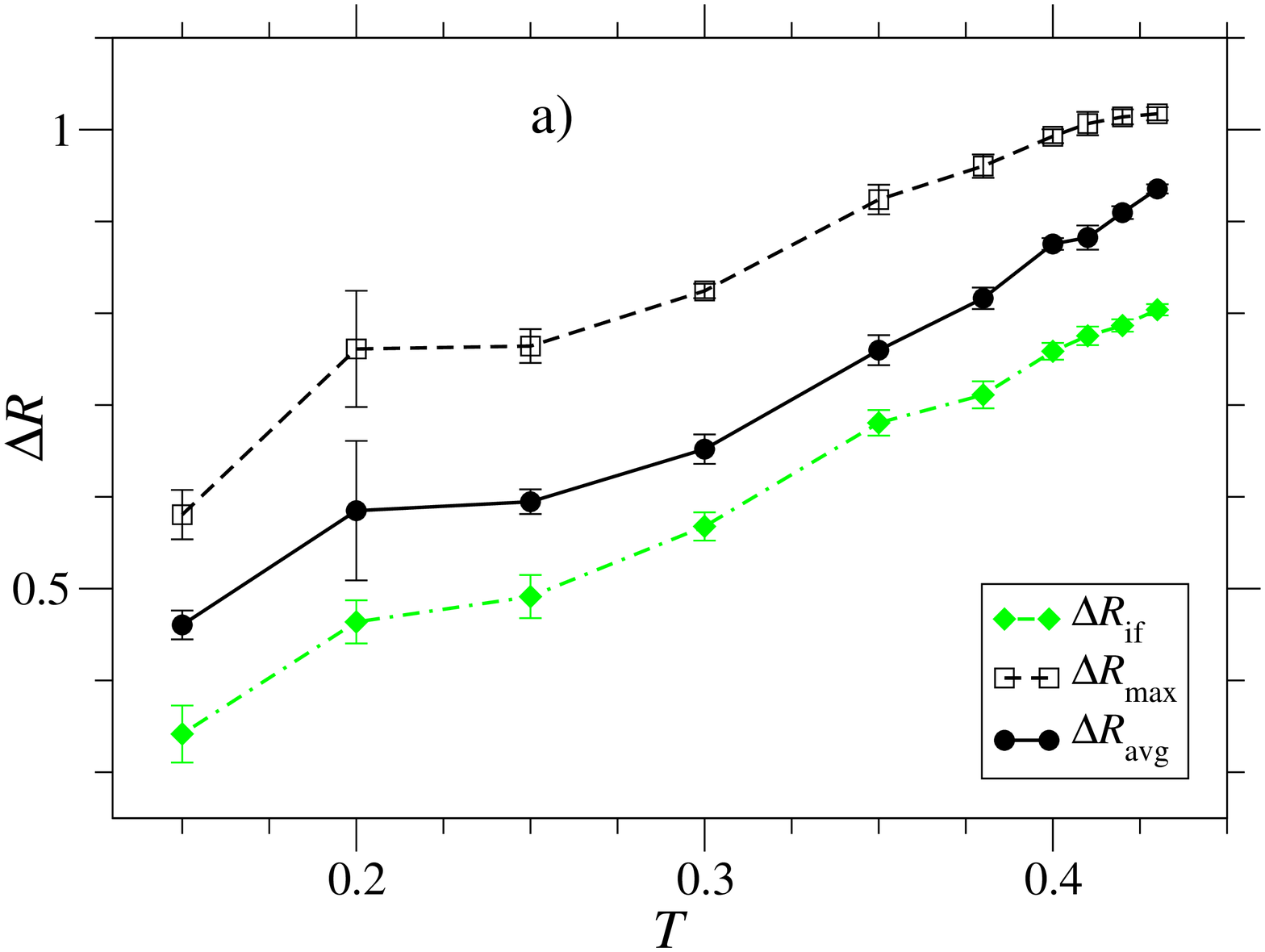}
\epsfxsize=3.2in
\epsfbox{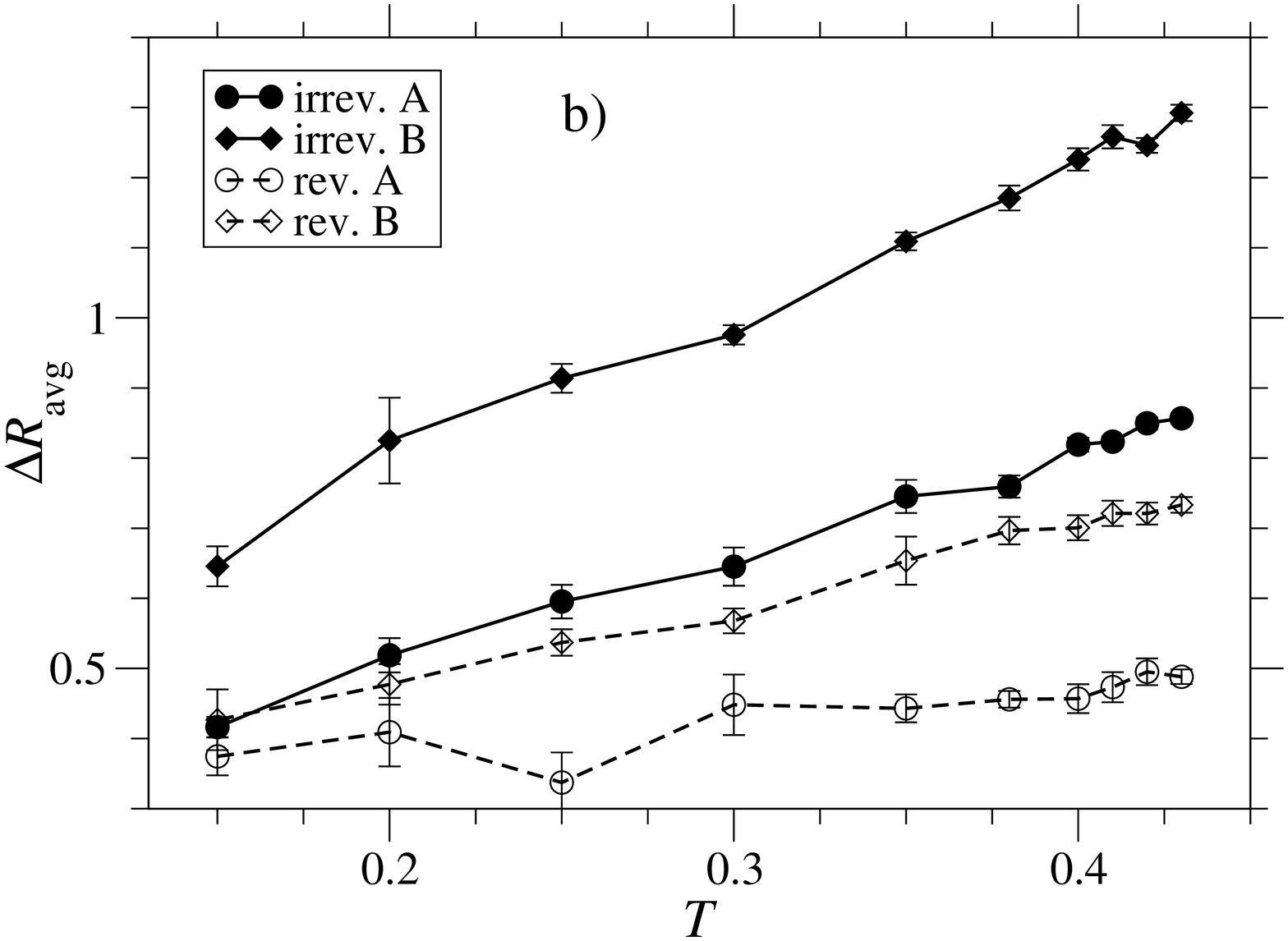}
\vspace*{2mm}
\caption{a) The jump sizes $\Delta R_\ind{if}$, $\Delta R_\ind{max}$
 and $\Delta R_\ind{avg}$ (sketched in Fig.~\ref{fig:dRifdRmaxdRba_def} and
 defined in the text) as function of temperature $T$. The averages 
 are over both irreversible and reversible jumps of both A and B particles.
 b) The jump size $\Delta R_\ind{avg}$ as function of temperature $T$
separately for irreversible and reversible 
jumps of A and B particles.
}
\label{fig:dRifmaxba_dRba}
\end{figure}
Fig.~\ref{fig:dRifmaxba_dRba}a
shows that the different jump lengths have the same qualitative 
behavior, namely, as one might expect, an increase with increasing 
temperature $T$. We therefore show in the following mainly 
results for only $\Delta R_\ind{avg}$ but find very similar behavior in
the case of $\Delta R_\ind{if}$ and $\Delta R_\ind{max}$.
An increase in jump size with increasing temperature has been seen in
previous simulations 
\cite{ref:Oligschleger660-811,ref:KlugePhD,ref:Miyagawa8278,ref:Schober67,ref:Oligschleger1031}.
However in the case of 
\cite{ref:Oligschleger660-811,ref:Schober67,ref:Oligschleger1031}
the jump size is for a jump in 
the MSD, i.e. an average over all particles, which includes two effects:
an increasing number of jumping particles 
and an increase in the jump size of single particle jumps.
Our approach has the advantage of keeping these two effects separate.
Furthermore, in our case the single particle jumps are larger
than in
\cite{ref:Oligschleger660-811,ref:Miyagawa8278,ref:Schober67,ref:Oligschleger1031}.

Fig.~\ref{fig:dRifmaxba_dRba}b is the same as Fig.~\ref{fig:dRifmaxba_dRba}a 
for $\Delta R_\ind{avg}$ but broken down by both  jump and particle
type.  The smaller 
B particles are jumping further than the A particles. Furthermore 
reversible jumps are shorter than the irreversible jumps, 
which is due to two features in the distribution 
$P(\Delta R_\ind{avg})$. 

\begin{figure}[htb]
\vspace*{1mm}
\epsfxsize=3.2in
\epsfbox{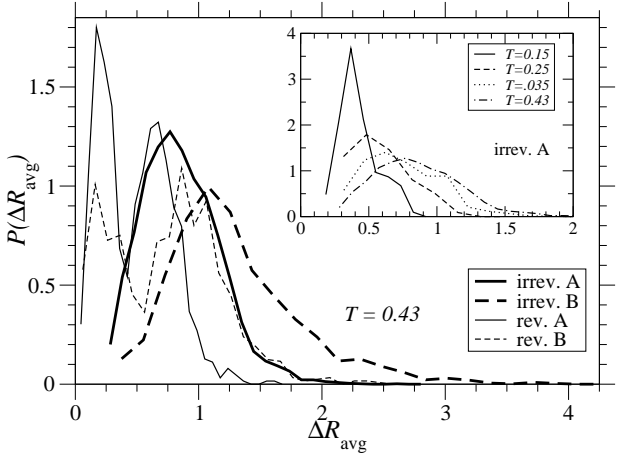}
\vspace*{2mm}
\caption{The distribution of jump sizes $\Delta R_\ind{avg}$ for irreversible 
and reversible jumps of A and B particles at $T = 0.43$. 
The inset shows $\Delta R_\ind{avg}$ of irreversible jumping A particles 
for different temperatures.
}
\label{fig:PofdRba_T430}
\end{figure}

As illustrated in
Fig.~\ref{fig:PofdRba_T430}  both irreversible and reversible jumps
have a peak at about $0.8$ and $1.0$ for A and B particles respectively.
These peaks are relatively
similar for reversible and irreversible jumps, taking aside that
for irreversible jumps this peak position is slightly shifted
to the right  and slightly broadened (partly due to
multiple jumps which are not resolved in time).
This similarity and the peak positions around unity are consistent 
with our picture that irreversible and reversible jumps start out
the same, namely with a particle jumping out of its cage.
A new feature, however, is that the distribution of reversible 
jumps for $T=0.43$ (and similarly for all other temperatures) has
a bimodal distribution with an additional 
peak at a distance smaller than $0.2$, 
contrary to the distribution of irreversible jumps.  
This seems to indicate for reversible jumps another process  than a
jump out of a cage because the latter  would be of the order 
$0.5$ and larger.
 
The increase of $\Delta R_\ind{avg}$ with increasing temperature 
is a consequence of both a shift of the peak position and a broadening 
of the distribution of jump sizes. This  is illustrated in the inset of 
Fig.~\ref{fig:PofdRba_T430} for irreversible jumps of A particles 
and we find similar distributions for irreversible jumps of B particles.
At the largest investigated temperature $T=0.43$ some of the 
particles move as far as four particle spacings.
Some of these large jumps might correspond to
multiple jumps of a smaller time window than our time resolution.
\clearpage

\begin{figure}[htb]
\epsfxsize=3.2in
\epsfbox{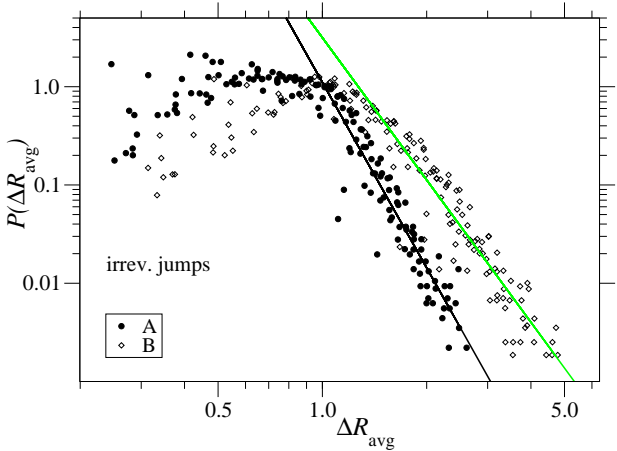}
\vspace*{2mm}
\caption{Log-log plot of the distribution of jump sizes $\Delta R_\ind{avg}$ 
for irreversible jumps of A and B particles including all temperatures.
The lines are linear fits with slopes $-6.3$ for A and $-4.8$ for 
B particles.
}
\label{fig:lnPoflndR}
\end{figure}
For irreversible jumps of size unity and larger we find that 
the distribution for all temperatures  follow roughly a power law 
$P(\Delta R) \approx \Delta R^{-\mu}$ (see Fig.~\ref{fig:lnPoflndR}) 
with $\mu = 6.3$ for A and $\mu = 4.8$ for B particles. This is in 
accordance with subdiffusive behavior for intermediate times
as presented in Sec.~\ref{sec:times} for the distribution of waiting 
times \cite{ref:eweeks_subdiff}.

As commented on earlier (see Sec.~\ref{sec:intro} and 
Sec~\ref{sec:jumpdef}),
our definition for a jump is not based on a specified size
but instead a multiple of the fluctuations of the 
particle.  Let us therefore next look at the fluctuations.
\begin{figure}[htb]
\epsfxsize=3.2in
\epsfbox{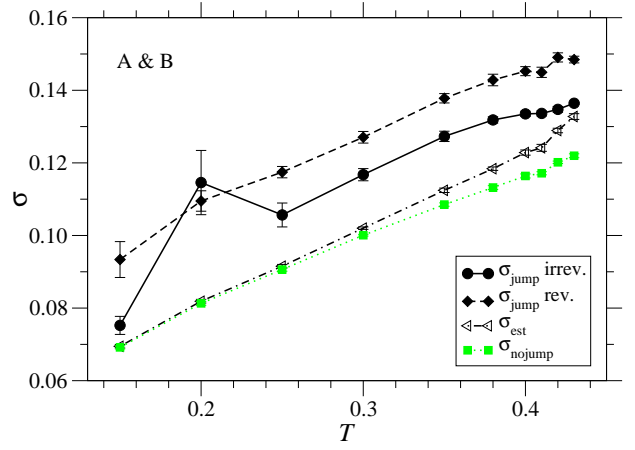}
\vspace*{3mm}
\caption{Fluctuations in position $\sigma$ for jumping particles 
  $\sigma_\ind{jump}$, of the original estimate for jumping particles
 $\sigma_\ind{est}$, and for non-jumping particles $\sigma_\ind{nojump}$.
  For the average of $\sigma_\ind{jump}$ fluctuations during
 the jump have been excluded from the average.
}
\label{fig:flucpos}
\end{figure}
Fig.~\ref{fig:flucpos} shows a comparison of the fluctuations of 
jumping particles $\sigma_\ind{jump}$ and of non-jumping particles 
$\sigma_\ind{nojump}$. Also included in Fig.~\ref{fig:flucpos} 
is an average over all particles $\sigma_\ind{est}$ of the estimates
for fluctuations $\sigma_\ind{i,est}$ as they have been used for 
the jump identification (see Sec.~\ref{sec:jumpdef} and
\cite{ref:refnoteIII1}). $\sigma_\ind{est}$ is very similar to 
$\sigma_\ind{nojump}$ and represents the fluctuations of an average
particle.
For the average of $\sigma_\ind{jump}$ 
we exclude fluctuations during the jump: 
for jumping particle $i$  we take
\begin{equation}
\sigma_\ind{i,jump} = \left ( \frac{
       \left ( \left \langle \sigma_i^2 \right \rangle_\ind{f} 
 +  \left \langle \sigma_i^2 \right \rangle_\ind{i} \right )
                                    }{2} \right )^{1/2}
\end{equation}
where $\sigma_i$ and $\langle \cdot \rangle_\ind{i,f}$ are defined as 
in \cite{ref:refnoteIII1} and earlier in this section respectively.
Although jump times are excluded, the fluctuations of jumping particles 
are clearly larger than the fluctuations of non-jumping particles. This
means that jumping particles are not only moving farther than an
average particle during single events but are also oscillating within
their cages with a larger than average amplitude.

\begin{figure}[htb]
\epsfxsize=3.2in
\epsfbox{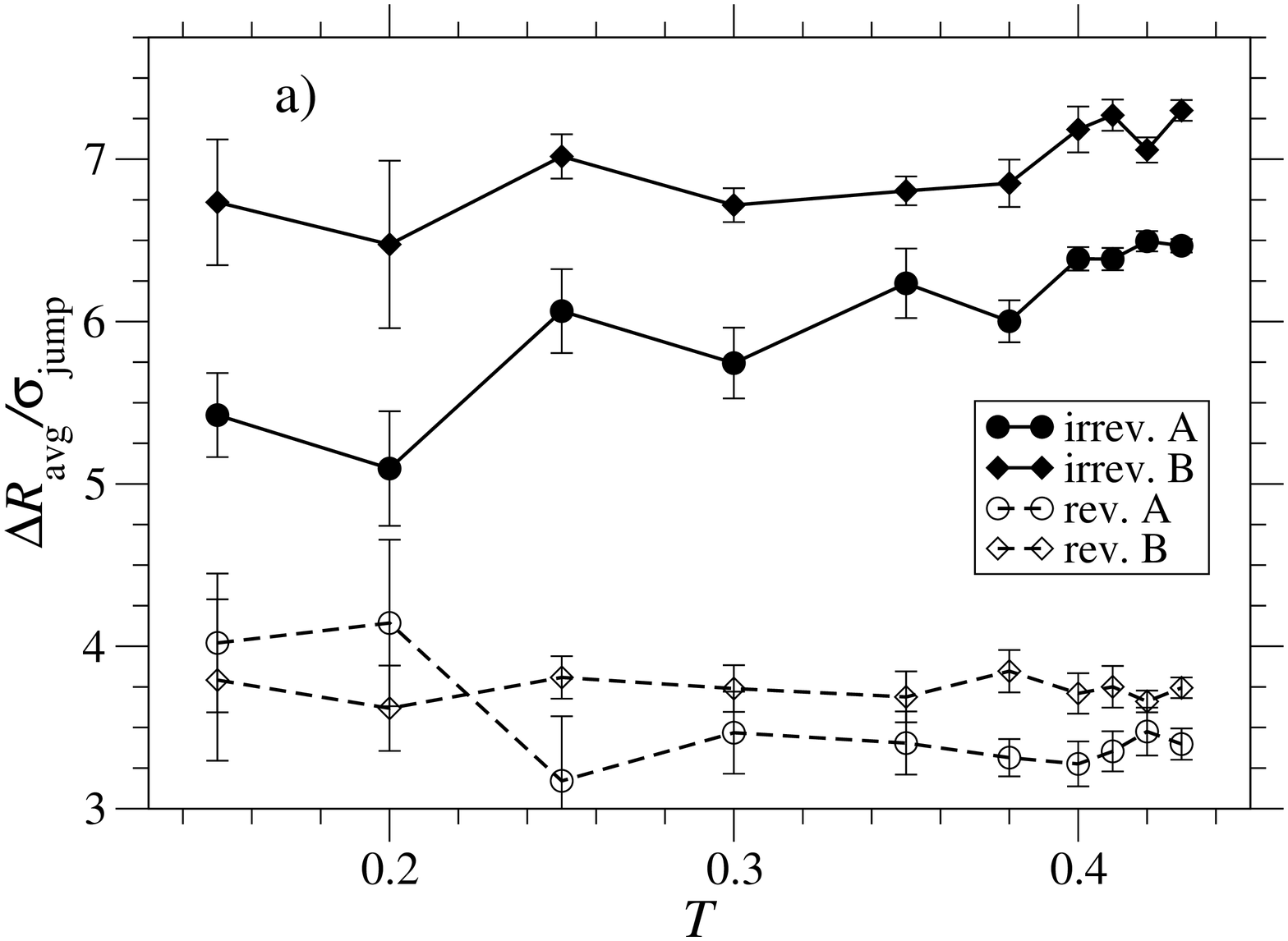}
\epsfxsize=3.2in
\epsfbox{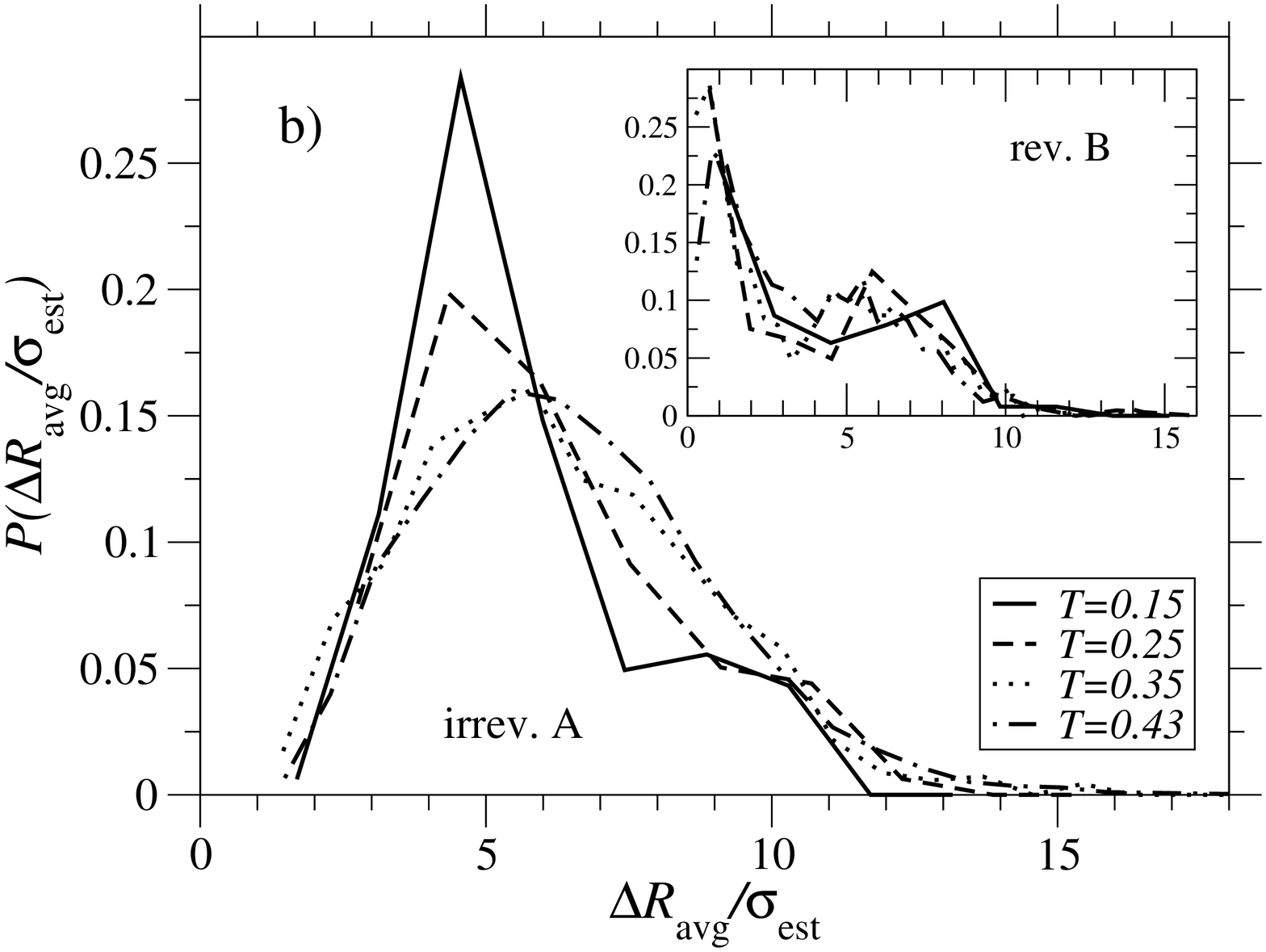}
\caption{a) Normalized jump size $\Delta R_\ind{avg}/\sigma_\ind{jump}$ 
  as function of temperature separately for irreversible and reversible jumps of
  A and B particles.
b) Distribution of normalized jump size 
$P(\Delta R_\ind{avg}/\sigma_\ind{est})$ of irreversible jumping 
A particles and reversible jumping B particles in the inset.
}
\label{fig:dRbadivsigba_PdRbadivsigflucest}
\end{figure}

\medskip
Fig.~\ref{fig:dRbadivsigba_PdRbadivsigflucest}a shows an average 
of jump size relative to 
the particle's fluctuation $\Delta R_\ind{i,avg}/\sigma_\ind{i,jump}$
for jumping particles $i$.
$\Delta R_\ind{avg}/\sigma_\ind{jump}$ seems independent of temperature 
for reversible jumps and for irreversible jumps slightly increasing with
increasing temperature.

\clearpage
\medskip
The distribution of $\Delta R/\sigma$ gives us the opportunity to
estimate the influence of the cutoff in our definition
of a jump. Similar to our approach for the search of a jump,
we use $\sigma_\ind{est}$, and obtain 
$P(\Delta R_\ind{avg}/\sigma_\ind{est})$ as shown in
Fig.~\ref{fig:dRbadivsigba_PdRbadivsigflucest}b 
for irreversible jumps of A particles. We
find a peak at 
$\left ( \Delta R_\ind{avg}/\sigma_\ind{est} \right ) \approx 5.5$
(for B-particles $\approx 6.5$), i.e. larger than our cutoff at 
$\left (\Delta R/\sigma \right ) = \sqrt{20} \approx 4.5$. 
This gives us the hope 
that we have included the major contribution of jumps consistent with
our approach. We have to bear in mind, however, that 
$\Delta R_\ind{avg}/\sigma_\ind{est}$ is not the identical measure 
to the one used in our search procedure.
Furthermore, in the case of reversible jumps there is some
additional peak at smaller values of $\left ( \Delta
R_\ind{avg}/\sigma_\ind{est} \right )$ (see inset of
Fig.~\ref{fig:dRbadivsigba_PdRbadivsigflucest}b) which is consistent with 
Fig.~\ref{fig:PofdRba_T430} and might indicate some additional 
process.

\section{Energy} \label{sec:energy}

In the previous sections we have investigated relaxation processes by 
solely using the positions (trajectories) of single particles. As 
the work 
\cite{ref:Sastry301,ref:DoliwaS849,ref:Schroeder9834,ref:Doliwa031506-Doliwa0306343,ref:Doliwa030501,ref:SaksaengwijitS1237}
shows, an approach via potential energies can also be 
fruitful. We therefore study in this section what happens to the
potential energy during jumps. 
\begin{figure}[htb]
\epsfxsize=3.2in
\epsfbox{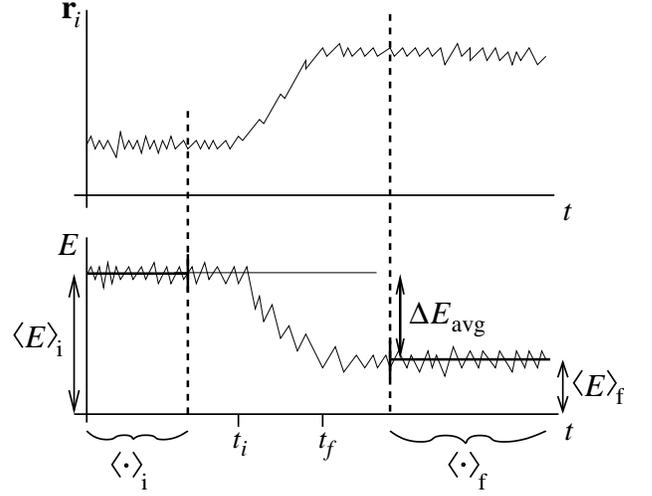}
\caption{Sketch for the definitions of $\langle E \rangle_\ind{i}$ 
and $\langle E \rangle_\ind{f}$.
}
\label{fig:dEba_def}
\end{figure}
As sketched in Fig.~\ref{fig:dEba_def}, 
we still use single particle trajectories to identify the jumping
particle and do similar averaging for 
the according time windows $\langle \cdot \rangle_\ind{i}$ 
and $\langle \cdot \rangle_\ind{f}$
(see Sec.~\ref{sec:jumplengths}). We investigate both the total potential
energy  per particle
\begin{equation}
\label{eq:Eoft}
E(t) = \frac{1}{N} \, \sum_{i=1}^N \sum_{j > i}
     V_{\alpha \beta} (r_{ij}(t))
\end{equation}
and the single particle potential energy  of particle $i$
\begin{equation}
\label{eq:Eioft}
E_i(t) = \sum_{j \neq i} V_{\alpha \beta} (r_{ij}(t))
\end{equation}
where $V_{\alpha \beta} (r_{ij}(t))$ is defined in 
Eq.~(\ref{eq:Valphabeta}).
To obtain in addition the energies of the inherent structures
\cite{ref:intrinsic_structure}
we minimized the configurations ($\{{\bf r}_i(t)\}$) via a steepest
decent procedure. With the thus obtained
configuration ($\{{\bf r}^0_i(t)\}$) we determine 
\begin{equation}
\label{eq:Eminoft}
E^0(t) = \frac{1}{N} \, \sum_{i=1}^N \sum_{j > i}
     V_{\alpha \beta} (r^0_{ij}(t))
\end{equation}
and
\begin{equation}
\label{eq:Eiminoft}
E^0_i(t) = \sum_{j \neq i} V_{\alpha \beta} (r^0_{ij}(t))
\end{equation}
Similar to the jump size in position 
(see Fig.~\ref{fig:dEba_def} and Sec.~\ref{sec:jumplengths}) we then 
determine 
\begin{equation}
\label{eq:dEba}
\Delta E_\ind{avg} = \langle E(t) \rangle_\ind{f} 
                    - \langle E(t) \rangle_\ind{i}
\end{equation}
and 
\begin{equation}
\label{eq:dEif}
\Delta E_\ind{if} = E(t_\ind{f}) - E(t_\ind{i})
\end{equation}
\clearpage
and similarly for the energies of Eq.~(\ref{eq:Eioft}) --
(\ref{eq:Eiminoft}).

We find in most of our analysis that $\Delta E_\ind{avg}$ and 
$\Delta E_\ind{if}$ (and all other equivalents for $E^0, E_i$ and
$E^0_i$) are showing the same qualitative behavior, and 
therefore present in the following results for $\Delta E_\ind{avg}$ 
using Eq.~(\ref{eq:dEba}) and Eq.~(\ref{eq:Eoft}) -- (\ref{eq:Eiminoft}).

\begin{figure}[htb]
\epsfxsize=3.2in
\epsfbox{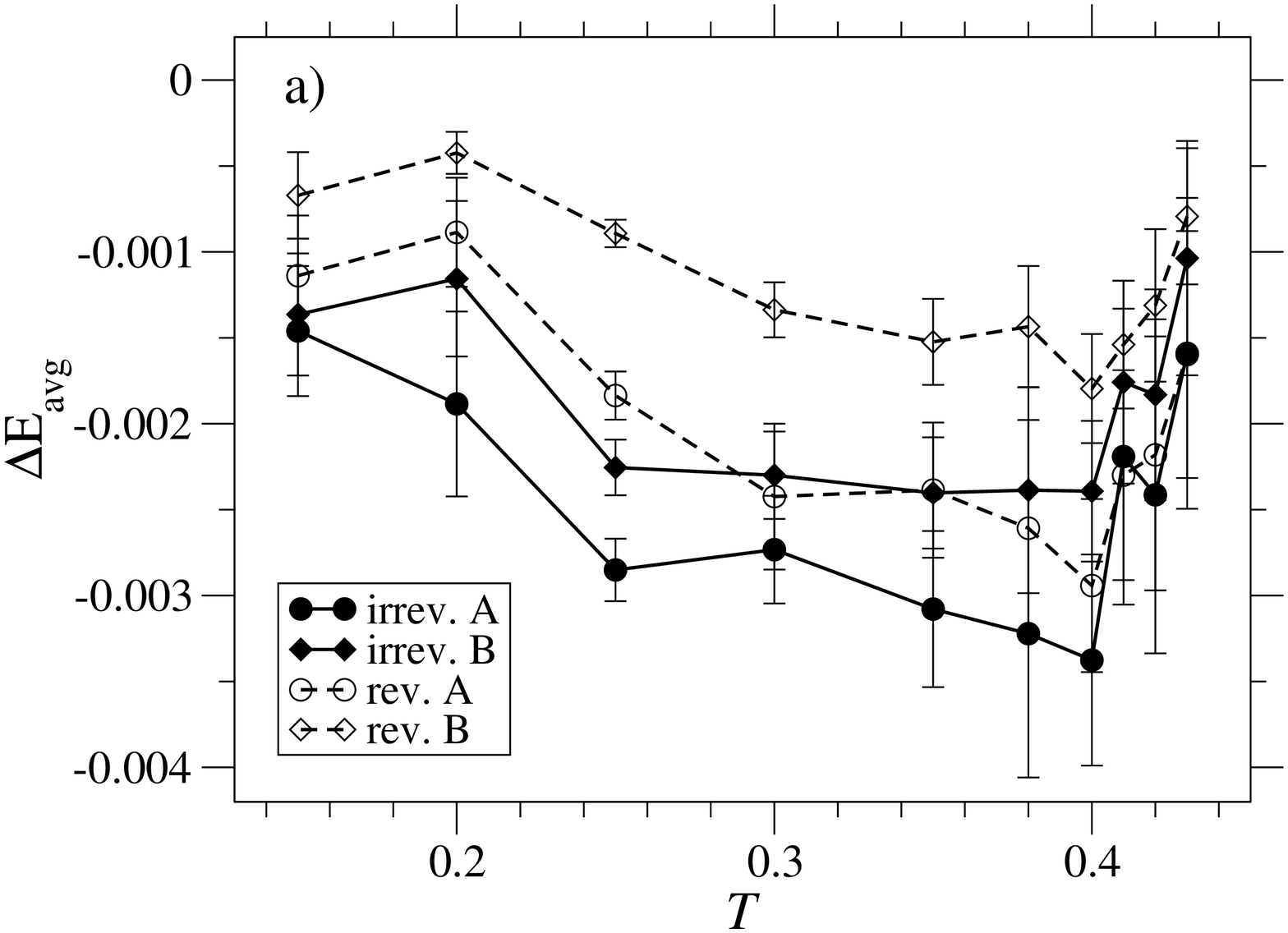}
\epsfxsize=3.2in
\epsfbox{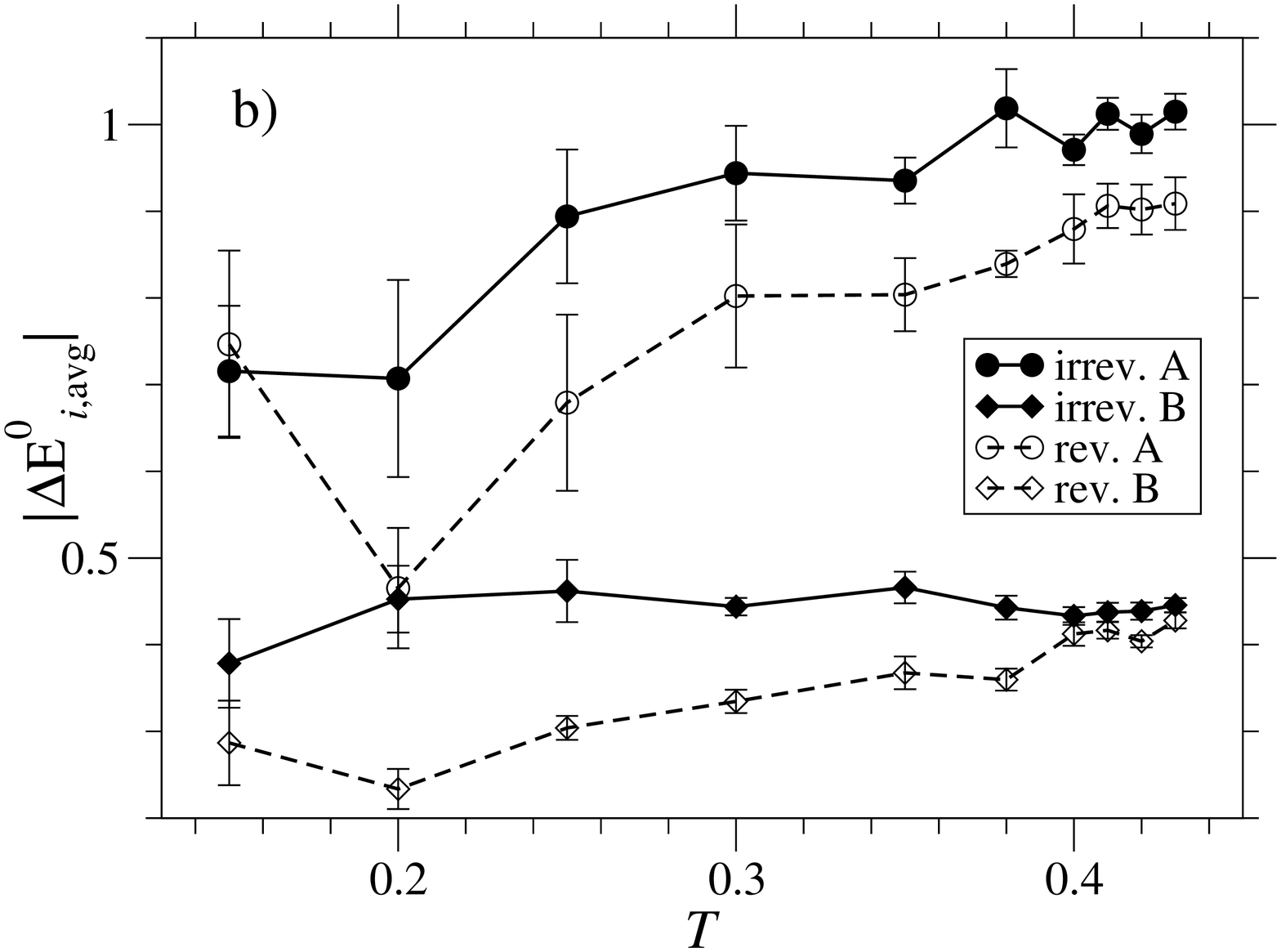}
\vspace*{2mm}
\caption{ As a function of temperature $T$
a) jump size in total potential energy $\Delta E_\ind{avg}$ 
and b)
absolute value of the jump size in minimized single particle 
potential energy $|\Delta E^{0}_{i\ind{,avg}}|$.
}
\label{fig:dEba_dEiminba}
\end{figure}
Fig.~\ref{fig:dEba_dEiminba}a shows $\Delta E_\ind{avg}$ as a function of
temperature averaged separately over irreversible and reversible jumps of A and
B particles. Notice that the energy jumps are small compared to
energy values $\left \langle E \right \rangle_\ind{i,f} \approx -7$. 
The irreversible jumps
lower the energies more than reversible jumps and more so for A than B
particles, because a jump of a larger A particle results in a bigger
change in environment than for a smaller B-particle. The lowering
of the total potential energy happens mostly at intermediate
temperatures and less at the lowest and highest investigated
temperatures. This might be an aging effect: 
at low temperatures the system
is basically frozen in, at intermediate temperatures the system has
time to partially age, and at higher temperatures the system has
already aged before the production run starts. This effect is lost
when we look at the jumps in single particle energies 
$\Delta E_{i,\ind{avg}}(T)$ which is
basically zero for all jumps. 
We therefore find that the total potential energies 
$\Delta E_\ind{avg}$, $\Delta E_\ind{if}$, $\Delta E^0_\ind{avg}$,
and $\Delta E^0_\ind{if}$ show a more systematic dependence on
temperature than their single particle equivalents 
$\Delta E_{i,\ind{avg}}$ etc.. For the absolute values 
$|\Delta E_\ind{avg}|$ etc. and all following quantities, however,
we find the opposite, i.e. the single particle equivalents 
$|\Delta E_{i,\ind{avg}}|$ etc. show a more pronounced behavior.
We therefore show from now on the
results which use Eq.~(\ref{eq:Eioft}) and Eq.~(\ref{eq:Eiminoft})
only. 

We next look at the absolute
value $|\Delta E^0_{i,\ind{avg}}|$ as a function of temperature (see
Fig.~\ref{fig:dEba_dEiminba}b). As one might expect, 
the irreversible jumps show larger changes in energy than reversible jumps, 
due to larger jumps in position (Fig.~\ref{fig:dRifmaxba_dRba}b) and
therefore more change in the  environment of the jumping particle. 
And similarly the larger A particles in comparison with the 
smaller B particles
experience a larger change in energy because A particles are surrounded
by more neighbors, i.e. a larger environment. 
With increasing temperature the absolute value of 
the jump in energy (Fig.~\ref{fig:dEba_dEiminba}b) is increasing (with
the one exception of irreversible B particles) which is consistent with the
increase in jump position (Fig.~\ref{fig:dRifmaxba_dRba}b).

\begin{figure}[htb]
\epsfxsize=3.2in
\epsfbox{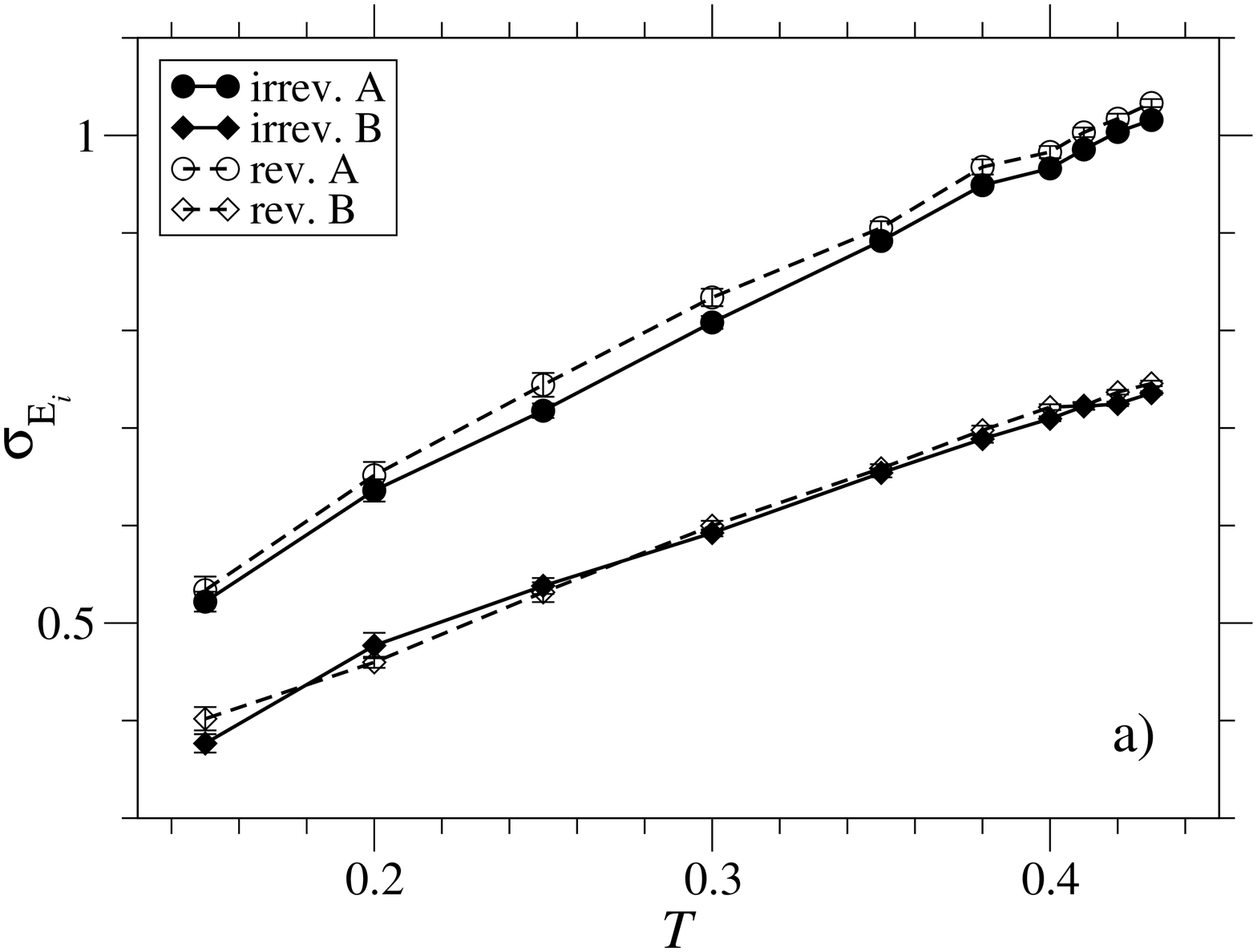}
\epsfxsize=3.2in
\epsfbox{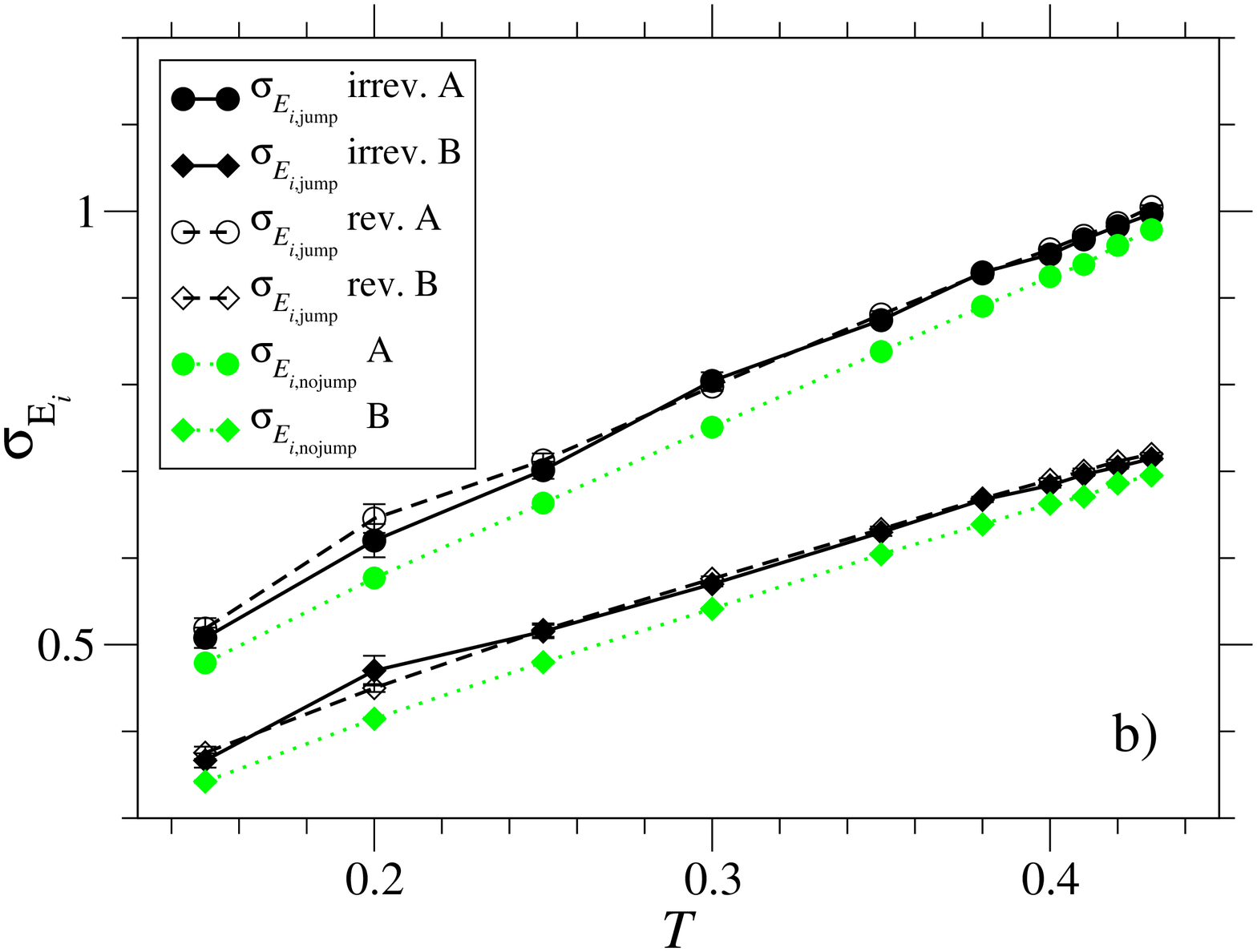}
\vspace*{2mm}
\caption{Fluctuations in single particle potential 
energy as a function of temperature.
The average is in a) over irreversible and reversible jump {\em events}
of A and B particles separately and in b) over
jumping {\em particles} ($\sigma_{E_{i,\ind{jump}}}$) and over
non-jumping {\em particles} ($\sigma_{E_{i,\ind{jump}}}$). 
}
\label{fig:sigEi_sigEinojump}
\end{figure}

We define the fluctuations in energy $\sigma_{E_i}$ at 
time $t=m \cdot 2000 \cdot 0.02$
\begin{eqnarray}
\sigma_{E_i} & = &  \left ( \frac{1}{2} \left \{
        \left \langle \frac{1}{20} 
             \sum \limits_{m'=m-9}^{m+10} \left ( E_i(m') -
              \overline{E}_i(m') \right )^2 \right \rangle_\ind{i} 
           \right . \right .
        \nonumber \\ 
  & & \quad  +  
        \left .  \left .
        \left \langle \frac{1}{20} 
             \sum \limits_{m'=m-9}^{m+10} \left ( E_i(m') -
              \overline{E}_i(m') \right )^2 \right \rangle_\ind{f}
        \right \}
        \right )^{-1/2}
\end{eqnarray}
similar to the fluctuations in position \cite{ref:refnoteIII1}
where $\overline{E}_i(m)$ is the time average 
$\frac{1}{20} \sum \limits_{m'=m-9}^{m+10} E_i(m')$.
The energy fluctuations are also increasing with increasing temperature
and are larger for A than B particles (see Fig.~\ref{fig:sigEi_sigEinojump}a).
The fluctuations of irreversible and reversible jumps are however very similar,
which is consistent with the picture that reversible jumps are ``failed''
irreversible jumps and in that sense start out the same way as irreversible jumps.

Fig.~\ref{fig:sigEi_sigEinojump}b is a comparison of energy fluctuations of
jumping (both irreversible and reversible and of both A and B) particles 
$\sigma_{E_{i\ind{,jump}}}$ and of non-jumping particles 
$\sigma_{E_{i\ind{,nojump}}}$ \cite{ref:refnoteVIII2}.
In $\sigma_{E_{i\ind{,jump}}}$ we exclude the times during the jump by 
using the time windows $\langle \cdot \rangle_\ind{i,f}$. We find that
the energy fluctuations of jumping particles are larger than for 
non-jumping particles which is consistent with the corresponding
fluctuations in position (see Fig.~\ref{fig:flucpos}).
\clearpage

\begin{figure}[htb]
\epsfxsize=3.2in
\epsfbox{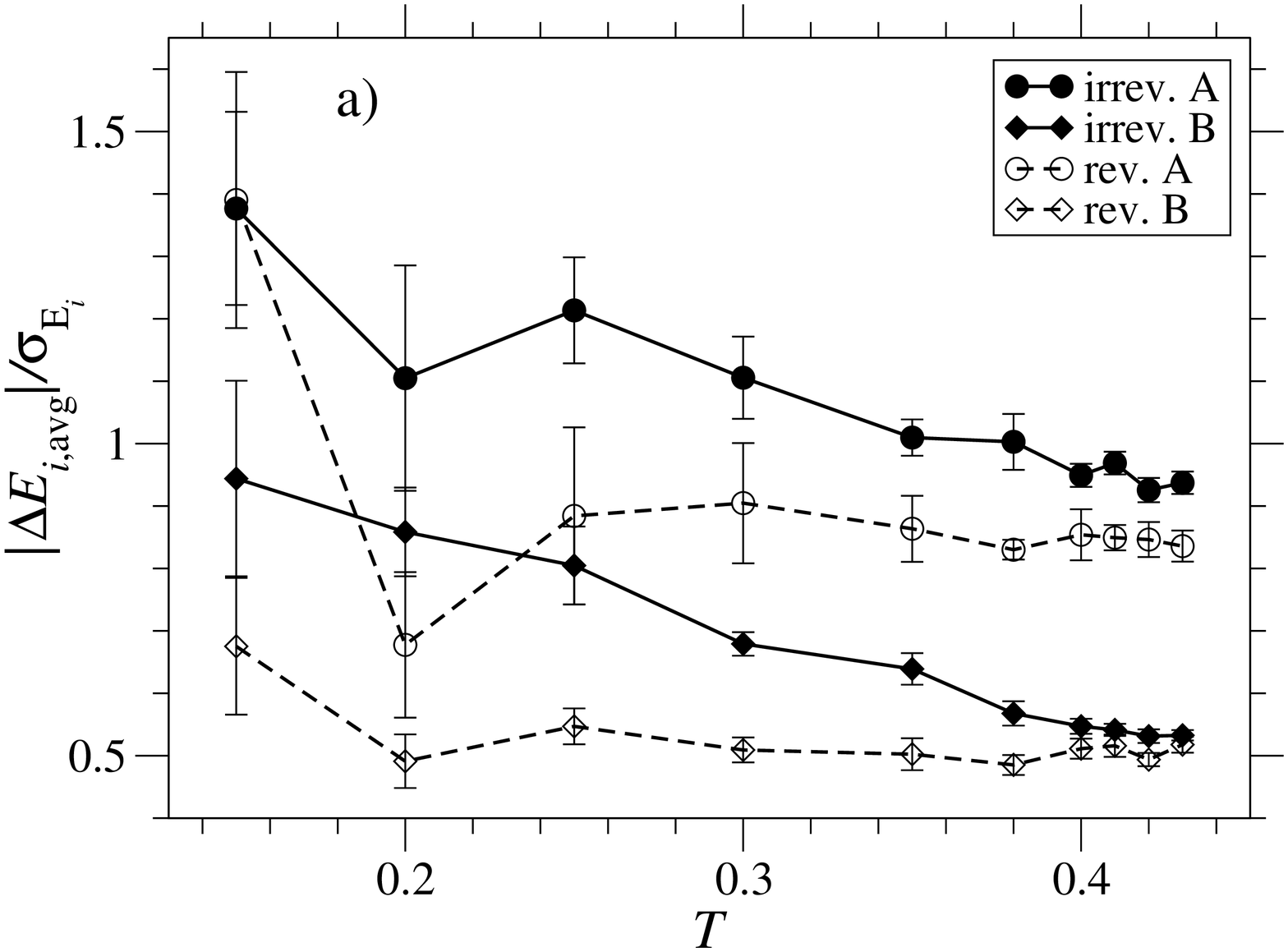}
\epsfxsize=3.2in
\epsfbox{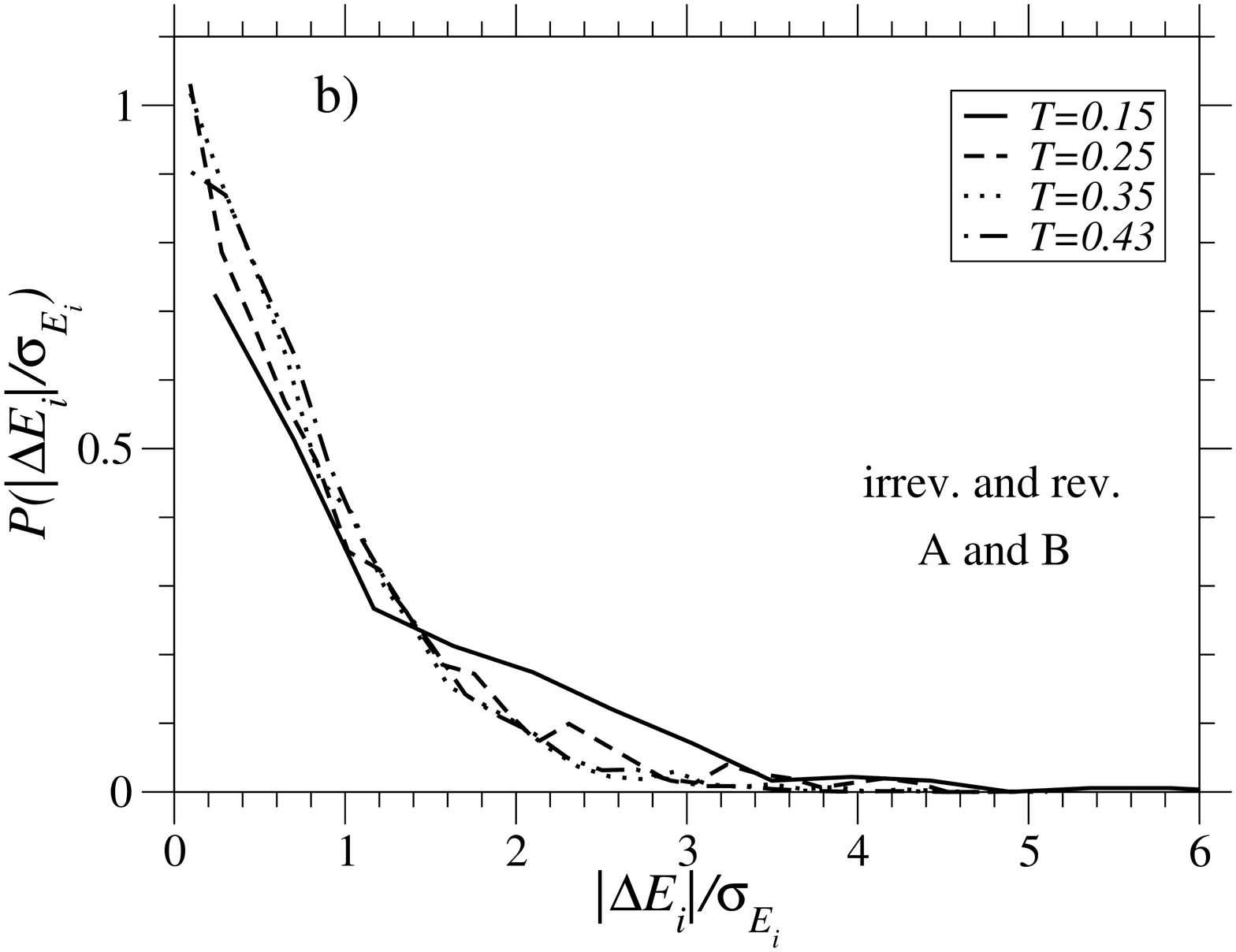}
\caption{Absolute value of single particle potential energy divided 
by its fluctuation $|\Delta E_{i,\ind{avg}}|/\sigma_{E_i}$  
in a) as a function of temperature and in b) its distribution .
}
\label{fig:dEidivsig_PofdEidivsig}
\end{figure}

By normalizing $|\Delta E_{i\ind{,avg}}|$ with the fluctuations
$\sigma_{E_i}$ we obtain Fig.~\ref{fig:dEidivsig_PofdEidivsig}a. Similar to the
case of normalized jumps in position 
(Fig.~\ref{fig:dRbadivsigba_PdRbadivsigflucest}a)
we find that $|\Delta E_{i\ind{,avg}}|/\sigma_{E_i}$ is for reversible
jumps basically independent of temperature. For irreversible jumps, however
$|\Delta E_{i\ind{,avg}}|/\sigma_{E_i}$ decreases significantly with
increasing temperature. This indicates that with increasing temperature
the jumps are increasingly more driven by fluctuations.
Contrary to the equivalent in position
(Fig.~\ref{fig:dRbadivsigba_PdRbadivsigflucest}a), the normalized energies are
significantly smaller, in the range 0.5 -- 1.4 rather than 3 -- 7,
and the distribution of $|\Delta E_{i\ind{,avg}}|/\sigma_{E_i}$ 
(see Fig.~\ref{fig:dEidivsig_PofdEidivsig}b)
is monotonous decreasing with a less far reaching tail 
than the equivalent in position 
(Fig.~\ref{fig:dRbadivsigba_PdRbadivsigflucest}b).

\section{Conclusions}
\label{sec:conclusions}

We study the dynamics of a binary Lennard-Jones system below the 
glass transition. Our focus is on jump-processes, which we identify
via single particle trajectories. Two kind of jumps are found:
``reversible jumps,'' where a particle jumps back and forth between
one or more average positions, and ``irreversible jumps,'' where
a particle does not return to any of its former average positions, i.e. 
successfully escapes its cage of neighbors. Both irreversible and reversible jumps
of A and B particles occur at all temperatures. With increasing
temperature more particles jump, more average positions are visited, the
jump size both in position and in the absolute value of the potential
energies $|\Delta E|, |\Delta E^0|, |\Delta E_i|$ and $|\Delta
E^0_i|$ (total and single-particle, not minimized and minimized) 
increase, and the fluctuations in position $\sigma_R$ and 
potential energy $\sigma_E$ increase. The fluctuations are larger
for jumping particles than for non-jumping particles even if
fluctuations during the jump are not part of the average, which
indicates that jumping particles are not only during jump times more
mobile than non-jumping particles. The ratio 
$|\Delta E_i|/\sigma_{E_i}$ of irreversible jumps decreases with increasing
temperature. This confirms a commonly used assumption that with
increasing temperature the irreversible jumps become more driven by
fluctuations. 

With increasing temperature also irreversible jumps become
proportionally more frequent
than reversible jumps. We interpret this such that irreversible
and reversible jumps are similar in that sense that a particle tries
to escape its cage. In the case of reversible jumps the particle finds
its way
back into the cage whereas in the case of an irreversible jump the path back
into the cage becomes blocked due to rearrangements of the cage. At
larger temperature these rearrangements of the cage become more
likely (since for example the fluctuations increase) and therefore
irreversible jumps occur more often. Irreversible and reversible jumps show in most
quantities qualitatively the same behavior, such as their
temperature dependence of jump size in position and energy, and
differ only in size. An exception to this similarity is 
the distribution of jump sizes, which
suggests that there might be an additional jump process in the case
of reversible jumps indicated by an additional peak at small jump
sizes.

The most surprising result of our work is that the times between
successive jumps are independent of temperature. 
This is most likely due to aging, which
could mean on the time scale of our simulation that 
$\Delta t_\ind{b}$ reflects the time scale of $T = 0.446$ from which
we quenched the system, and that $\Delta t_\ind{b}$ might show
temperature dependence at later times. The latter would be
consistent with the work of Doliwa {\it et al.} 
\cite{ref:Doliwa031506-Doliwa0306343,ref:Doliwa030501} who
find in their simulations
of very long times that the distribution of waiting times 
(however, identified via the collective quantity of the minimized
potential energy) initially is temperature independent and of
different power law than at later times when temperature dependence occurs. 
We do find aging in that sense that
the times before the first jump are shorter than
the times after the last jump of a particle (an effect which becomes
less for temperatures near $T_c$). 
Future work (which is in progress) will tell us if in
our system $\Delta t_\ind{b}$ becomes temperature dependent after
longer times.
However, the dynamics does depend on temperature for all other here 
investigated quantities such as the number of jumps and the fraction of 
reversible jumps.

Another question which we would like to raise for future work is
how much single particle jumps can tell us about relaxations of a
glass which might include collective motion. It seems plausible that
single particle jumps are strongly correlated with collective jumps
and in that sense give us very similar information to studies 
of collective quantities. We find that many of the single particle 
jumps in our system are spatially and temporally correlated, i.e. 
are showing collective motion. Systematic studies for confirmation
remain to be done in the future. We also find the signature of 
many-particle effects confirmed in the result of the increasing 
ratio of irreversible jumps with increasing temperature, since this
indicates that not only the jumping particle itself but also its 
cage are dependent on temperature. 
There might, however, be also collective jumps, where each particle
jumps a too small amount to be detected by our search algorithm. It
remains to be studied of how much importance such processes are in
our system.

\acknowledgments

I would like to thank  K.~Binder and  A.~Zippelius for
hospitality, financial support and for their helpful discussions. 
I also thank J.~Horbach for helpful discussions and B.~Vollmayr-Lee
and J.~Horbach for the careful 
reading of this manuscript.
I gratefully acknowledge financial support from 
SFB 262 and DFG Grant No. Zi 209/6-1.



\begin{references}
\bibitem{ref:glassintro}  For reviews see e.g.
  R.~Zallen,  {\it The Physics of Amorphous Materials} (Wiley, New York, 1983);
  J.~J\"ackle, Rep.\ Prog.\ Phys. {\bf 49}, 171 (1986);
  W.~G\"otze and L.~Sj\"ogren, Rep.\ Prog.\ Phys. {\bf 55}, 241 (1992);
  C.\ A.~Angell, Science {\bf 267}, 1924 (1995); 
  Proceedings of  3rd International Discussion Meeting on Relaxation
    in Complex Systems Ed.: K.\ L. Ngai, J.\ Non-Cryst.\ Solids {\bf
    235-238} (1998).
\bibitem{ref:MSDexamples} Examples 
  of the mean square displacement as a function of time 
  (logarithmically plotted) are given in
   W.~Kob and H.~C. Andersen, Phys.\ Rev.\ E {\bf 51}, 4626 (1995). 
\bibitem{ref:Oligschleger660-811} C. Oligschleger, C. Gaukel and 
   H.~R.~Schober,  J.~Non-Cryst.~Solids {\bf 250-252}, 660 (1999);
   C.~Oligschleger and H.~R.~Schober, Phys.\ Rev.\ B {\bf 59}, 811 (1999).
\bibitem{ref:Teichler339-533} H. Teichler,  J.~Non-Cryst.~Solids 
   {\bf 293-295}, 339 (2001);
   H. Teichler, J.~Non-Cryst.~Solids {\bf 312-314}, 533 (2002).
\bibitem{ref:Schober40} H.~R.~Schober,  J.~Non-Cryst.~Solids
 {\bf 307}, 40 (2002).
\bibitem{ref:Gaukel1907} C. Gaukel, M. Kluge and H.~R.~Schober,
  Phil.~Mag.~B {\bf 79}, 1907 (1999).
\bibitem{ref:Caprion3709} D. Caprion and H.~R.~Schober,
   Phys.\ Rev.~B {\bf 62}, 3709 (2000).
\bibitem{ref:Schober723} H.~R.~Schober, C.~Gaukel, and
  C.~Oligschleger,  Defect~and~Diffusion~Forum {\bf 143}, 723 (1997). 
\bibitem{ref:Sanyal4154} S.~Sanyal and A.~K.~Sood, 
      Phys.\ Rev.\ E {\bf 52}, 4154 (1995).
\bibitem{ref:Delaye2063} J.~M. Delaye and Y. Limoge,  J.\ Phys.\ I
   {\bf 3}, 2063 (1993).
\bibitem{ref:Lacevic030101-Glotzer509} N. La\v{c}evi\'{c}, F.~W.~Starr, 
   T.~B.~Schr{\o}der, V.~N.~Novikov, and S.~C.~Glotzer,
    Phys.\ Rev.\ E {\bf 66}, 030101(R) (2002);
   S.~C.~Glotzer, V.~N.~Novikov, and 
   T.~B.~Schr{\o}der,  J.\ Chem.\ Phys. {\bf 112}, 509 (2000).
\bibitem{ref:Glotzer214} S.~C.~Glotzer, Y. Gebremichael,
   N. La\v{c}evi\'{c}, T.~B.~Schr{\o}der, and F.~W.~Starr,
    ACS~Sym.~Ser. {\bf 820}, 214 (2002).
\bibitem{ref:DoliwaA277} B. Doliwa and A. Heuer, 
    J.~Phys.:Condens.~Matter {\bf 11}, A277 (1999). 
\bibitem{ref:Doliwa4915} B. Doliwa and A. Heuer, 
    Phys.\ Rev.\ Lett. {\bf 80}, 4915 (1998). 
\bibitem{ref:Roe1610} R.-J. Roe,  J.\ Chem.\ Phys. {\bf 100}, 1610 (1994). 
\bibitem{ref:Sanyal361} S. Sanyal and A.~K.~Sood, 
           Europhys.~Lett. {\bf 34}, 361 (1996).
\bibitem{ref:Wahnstroem3752} G. Wahnstr\"om,  Phys.\ Rev.\ A 
    {\bf 44}, 3752 (1991). 
\bibitem{ref:Bhattacharyya2741} S.~Bhattacharyya, A.~Mukherjee, and B.~Bagchi,
    J.\ Chem.\ Phys. {\bf 117}, 2741 (2002). 
\bibitem{ref:KlugePhD} M. Kluge, Ph.D. thesis,  Berichte des
              Forschungszentrums J\"ulich {\bf 3913}, 
              J\"ulich, Germany (2001).
\bibitem{ref:Gaukel1} C. Gaukel and H.~R. Schober,  Sol. State Commun.
    {\bf 107}, 1 (1998). 
\bibitem{ref:Gaukel664} C. Gaukel, M. Kluge and H.~R.\ Schober, 
     J.~Non-Cryst.~Solids {\bf 250-252}, 664 (1999). 
\bibitem{ref:Schroeder331} T.~B.~Schr{\o}der, J.~C.~Dyre,
     J.~Non-Cryst.~Solids {\bf 235-237}, 331 (1998).
\bibitem{ref:Roux7171} J.~N. Roux, J.~L. Barrat and J.-P. Hansen,
    J.\ Phys.:Condens.\ Matter {\bf 1}, 7171 (1989). 
\bibitem{ref:Bruening964} R. Br\"uning, D.~H.~Ryan,
J.~O.~Str\"om-Olsen, and L.~J.~Lewis,  Material Science and 
   Engineering {\bf A134}, 964 (1991).
\bibitem{ref:Kob2827} W. Kob, C. Donati, S. J. Plimpton,
  P. H. Poole and S. C. Glotzer, Phys.\ Rev.\ Lett. {\bf 79}, 2827 (1997)
\bibitem{ref:Donati2338-Donati3107} 
   C. Donati, J. F. Douglas, W. Kob, S.J. Plimpton, 
  P.H. Poole and S.C. Glotzer, Phys.\ Rev.\ Lett. {\bf 80}, 2338 (1998);
  C. Donati, S.\ C. Glotzer, P.\ H. Poole, W. Kob
  and S.\ J. Plimpton, Phys.\ Rev.\ E {\bf 60}, 3107 (1999).
\bibitem{ref:Yamamoto4915} R.~Yamamoto and A.~Onuki,
  Phys.\ Rev.\ Lett. {\bf 81}, 4915 (1998).
\bibitem{ref:Kegel290} W.~K.~Kegel and A.~van~Blaaderen, 
   Science {\bf 287}, 290 (2000). 
\bibitem{ref:Weeks627} E.~R.~Weeks, J.~C.~Crocker, A.~C.~Levitt,
  A.~Schofield, and D.~A.~Weitz,  Science {\bf 287}, 627 (2000).
\bibitem{ref:Weeks095704} E.~R.~Weeks and D.~A.~Weitz, 
     Phys.\ Rev.\ Lett. {\bf 89}, 095704 (2002).
\bibitem{ref:Hurley10521} M.~M.~Hurley and P.~Harrowell,
      J.\ Chem.\ Phys. {\bf 105}, 10521 (1996).
\bibitem{ref:Caprion4293} D.~Caprion, J.~Matsui, and H.~R.~Schober,
   Phys.\ Rev.\ Lett. {\bf 85}, 4293 (2000). 
\bibitem{ref:RahmanA405} A. Rahman, Phys.\ Rev. {\bf 136}, A405 (1964).
\bibitem{ref:Miyagawa8278} H.~Miyagawa and Y.~Hiwatari, 
   Phys.\ Rev.\ A {\bf 44}, 8278 (1991).
\bibitem{ref:dynhetreviews} For reviews see 
    M.\ D. Ediger, Ann.\ Rev.\ Phys.\ Chem. {\bf 51}, 99 (2000); 
    H. Sillescu, J.\ Non-Cryst.\ Solids {\bf 243}, 81 (1999);
    R. B\"ohmer, Curr.\ Opin.\ Solid State\ Mat.\ Sci. {\bf 3}, 378 (1998).
\bibitem{ref:exp_dynhet}R. Richert, J.\ Non-Cryst.\ Solids {\bf
 172-174}, 209 (1994);
 R.\ Richert, J.\ Phys.\ Chem.\ {\bf 101}, 6323 (1997);
 F.\ R. Blackburn, M.\ T. Cicerone, G.\
 Hietpas, P.\ A.\ Wagner, M.\ D.\ Ediger, J.\ Non-Cryst.\ Solids {\bf
 172-174}, 256 (1994);
 K.\ Schmidt-Rohr and H.\ W.\ Spiess,
  Phys.\ Rev.\ Lett. {\bf 66}, 3020 (1991);
 J.\ Leisen, K.\ Schmidt-Rohr and H.\
 W.\ Spiess, J.\ Non-Cryst.\ Solid {\bf 172-174}, 737 (1994);
 A.\ Heuer, M.\ Wilhelm, H.\ Zimmermann and H.\ W.\ Spiess, Phys.\
 Rev.\ Lett.\ {\bf 75}, 2851 (1995);
 M.\ T.\ Cicerone, F.\ R.\ Blackburn and M.\ D.\ Ediger, J.\ Chem.\
 Phys.\ {\bf 102}, 471 (1995);
 M.\ T.\ Cicerone and M.\ D.\ Ediger, J.\ Chem.\ Phys.\ {\bf 103}, 5684
 (1995);
 F.\ Fujara, B.\ Geil, H.\ Sillescu and G.\ Fleischer, Z.\ Physik B {\bf
 88}, 195 (1992);
 T. Kanaya, U. Buchenau, S. Koizumi, I. Tsukushi and K. Kaji,
      Phys.\ Rev.\ B {\bf 61}, R6451 (2000);
 M. Russina, F. Mezei, R. Lechner, S. Longeville and B. Urban,
   Phys.\ Rev.\ Lett. {\bf 84}, 3630 (2000);
 W. Schmidt, M. Ohl and U. Buchenau, Phys.\ Rev.\ Lett. {\bf 85}, 5669
   (2000);
 L.\ F. Cugliandolo, J. Iguain, Phys.\ Rev.\ Lett.\ {\bf 85}, 3448 (2000);
 G. Diezemann, G. Hinze and H. Sillescu, 
  J.~Non-Cryst.~Solids {\bf 307}, 57 (2002);
 R.~E.~Courtland and E.~R.~Weeks,
    J.~Phys.:Condens.~Matter {\bf 15}, S359 (2003).
\bibitem{ref:MuranakaR2735-Gould5707-Perera120-Doliwa32}
   T. Muranaka and Y. Hiwatari,Phys.\ Rev.\ E {\bf 51}, R2735 (1995);
   G. Johnson, A.~I.~Mel'cuk, H. Gould, W. Klein and 
    R.\ D. Mountain, Phys.\ Rev.\ E {\bf 57}, 5707 (1998);
   D.~N.~Perera and P.~Harrowell, Phys.\ Rev.\ Lett. {\bf 81}, 120 (1998);
   B.\ Doliwa and A.\ Heuer, 
     J.~Non-Cryst.~Solids {\bf 307}, 32 (2002).
\bibitem{ref:Hurley1694} M.~M.~Hurley and P.~Harrowell,
    Phys.\ Rev.\ E {\bf 52}, 1694 (1995).
\bibitem{ref:Perera5441} D.~N.~Perera and P.~Harrowell,
    J.\ Chem.\ Phys. {\bf 111}, 5441 (1999).
\bibitem{ref:Doliwa6898} B.~Doliwa and A.~Heuer,
   Phys.\ Rev.\ E {\bf 61}, 6898 (2000).
\bibitem{ref:Gebremichael051503-Heuer6176-Yamamoto3515-Gaukel67-Poole51}
   A.~Heuer and K.~Okun, J.\ Chem.\ Phys.  {\bf 106}, 6176 (1997);
   H.\ R. Schober, C. Gaukel and C. Oligschleger,
   Prog.\ Theor.\ Phys.\ Suppl. {\bf 126}, 67 (1997);
  R.\ Yamamoto and A.\ Onuki, Phys.\ Rev.\ E {\bf 58}, 3515 (1998);
  P. H. Poole, C. Donati and S. C. Glotzer,  Physica A {\bf 261}, 51 (1998);
   Y. Gebremichael, T.~B.~Schr{\o}der,
   F.~W.~Starr, and S.~C.~Glotzer,  Phys.\ Rev.\ E {\bf 64}, 
   051503 (2001).
\bibitem{ref:kvldynhetJCP} K. Vollmayr-Lee, W. Kob, K. Binder and A.
    Zippelius, J.\ Chem.\ Phys. {\bf 116}, 5158 (2002).
\bibitem{ref:Teichler717} H. Teichler,
   Defect\ Diffus.\ Forum {\bf 143-147}, 717 (1997).
\bibitem{ref:Laird636} B.~B.~Laird and H.~R.~Schober, 
    Phys.\ Rev.\ Lett. {\bf 66}, 636 (1991).
\bibitem{ref:Schober6746} H.~R.~Schober and B.~B.~Laird, 
     Phys.\ Rev.\ B {\bf 44}, 6746 (1991).
\bibitem{ref:Schober965} H.~R.~Schober, C.~Oligschleger, and
  B.~B.~Laird,  J.~Non-Cryst.~Solids {\bf 156}, 965 (1993).
\bibitem{ref:CaprionMRS2002} D. Caprion, M. Kluge and H.~R.~Schober, 
   MRS Symposia Proceedings No.~754 (Materials Research Society,
    Pittsburg, 2003).  
\bibitem{ref:Schober67} H.~R.~Schober, C.~Gaukel, and
  C.~Oligschleger,  Progr.\ Theor.\ Phys.\ Suppl. {bf 126}, 67 (1997)
\bibitem{ref:Schober11469-Bembenek936-Keyes4651-Buchenau5039}
  U. Buchenau, Yu.~M.~Galperin, V.~L.~Gurevich, and H.~R.~Schober, 
      Phys.\ Rev.\ E {\bf 43}, 5039 (1991);
   S.~D.~Bembenek and B.~B.~Laird,
     Phys.\ Rev.\ Lett. {\bf 74}, 936 (1995);
   H.~R.~Schober and C.~Oligschleger,
    Phys.\ Rev.\ B {\bf 53}, 11469 (1996);
   T.~Keyes, G.~V.~Vijayadamodar, and
       U.~Zurcher,  J.\ Chem.\ Phys. {\bf 106}, 4651 (1997).
\bibitem{ref:ageing_review} For a review see W. Kob,
  ``Supercooled liquids and glasses,'' in  The Scottish
Universities Summer School, {\it Soft and Fragile Matter - Non-Equilibrium
Dynamics, Metastability and Flow}, Edingburgh, 2000.
\bibitem{ref:Bouchaud243} J.~-P.~Bouchaud, L.~F.~Cugliandolo,
   J.~Kurchan, and M.~M\'ezard, Physica A {\bf 226}, 243 (1996);
   J.~-P.~Bouchaud, L.~F.~Cugliandolo, J.~Kurchan, and M.~M\'ezard,  
   in {\it Spin Glasses and Random Fields}, Ed.: A.~P.~Young 
   (World Scientific, Singapore, 1998).
\bibitem{ref:Struik} L.~C.~E.~Struik, {\it Physical Aging in Amorphous 
   Polymers and Other Materials}, (Elsevier, Amsterdam, 1978).
\bibitem{ref:Kob4581} W.~Kob and J.~-L.~Barrat, Phys.\ Rev.\ Lett. 
    {\bf 78}, 4581 (1997).
\bibitem{ref:Barrat637} J.~-L.~Barrat and W.~Kob, Europhys.~Lett.
    {\bf 46}, 637 (1999).
\bibitem{ref:Sastry301} S. Sastry, P.~G.~Debenedetti,
  F.~H.~Stillinger, T.~B.~Schr{\o}der, J.~C.~Dyre, S.~C.~Glotzer,
  Physica A {\bf 270}, 301 (1999).
\bibitem{ref:DoliwaS849} B. Doliwa and A. Heuer,
  J.~Phys.:Condens.~Matter {\bf 15}, S849 (2003).
\bibitem{ref:Schroeder9834}
  T.\ B. Schr\o der, S. Sastry, J.\ C. Dyre and S.\ C. Glotzer,
   J.\ Chem.\ Phys. {\bf 112}, 9834 (2000).
\bibitem{ref:Doliwa031506-Doliwa0306343}
    B. Doliwa and A. Heuer, Phys.\ Rev.\ E {\bf 67}, 031506 (2003);
    B. Doliwa and A. Heuer, e-print cond-mat/0306343 (2003).
\bibitem{ref:Doliwa030501} 
    B. Doliwa and A. Heuer, Phys.\ Rev.\ E {\bf 67}, 
         030501(R) (2003). 
\bibitem{ref:SaksaengwijitS1237} A.~Saksaengwijit, B.~Doliwa, and
  A.~Heuer, J.~Phys.:Condens.~Matter {\bf 15}, S1237 (2003).
\bibitem{ref:Weeks361} E.~R.~Weeks and D.~A.~Weitz,
    Chem.\ Phys. {\bf 284}, 361 (2002).
\bibitem{ref:Rabani3649} E.~Rabani, J.~D.~Gezelter, and
  B.~J.~Berne,  Phys.\ Rev.\ Lett. {\bf 82}, 3649 (1999).  
\bibitem{ref:Rabani6867} E.~Rabani, J.~D.~Gezelter, and B.~J.~Berne,
  J.\ Chem.\ Phys. {\bf 107}, 6867 (1997).
\bibitem{ref:Gezelter3444} J.~D.~Gezelter, E.~Rabani, and B.~J.~Berne,
   J.\ Chem.\ Phys. {\bf 110}, 3444 (1999).
\bibitem{ref:Allegrini5714} P. Allegrini, J.~F.~Douglas, 
  and S.~C.~Glotzer, Phys.\ Rev.\ E {\bf 60}, 5714 (1999). 
\bibitem{ref:Leutheusser2765-Bengtzelius5915} 
     E.~Leutheusser, Phys.\ Rev.\ A {\bf 29}, 2765 (1984);
   U. Bengtzelius, W. G\"otze, and
     A.~Sj\"olander, J.\ Phys.\ C {\bf 17}, 5915 (1984).
\bibitem{ref:Goetze415-Das2265-Goetze3407-Sjoegren5-Fuchs7709} 
   W. G\"otze and L.~Sj\"ogren, Z.\ Phys.\ B {\bf 65}, 415 (1987);
   S.~P.~Das and G.~F.~Mazenko, Phys.\ Rev.\ A {\bf 34}, 2265 (1986);
    W. G\"otze and L.~Sj\"ogren, J.\ Phys.\ C {\bf 21}, 3407 (1988);
    L. Sj\"ogren, Z.\ Phys.\ B {\bf 79}, 5 (1990);
    M.~Fuchs, W.~G\"otze, S.~Hildebrand, and
   A.~Latz, J.~Phys.:Condens.~Matter {\bf 4}, 7709 (1992).
\bibitem{ref:Li43-Goetze4133} G.~Li, M.~Fuchs, W.~M.~Du, A.~Latz, 
    N.~J.~Tao, J.~Hernandez, W.~G\"otze, and H.~Z.~Cummins,
    J.~Non-Cryst.~Solids {\bf 172-174}, 43 (1994).
\bibitem{ref:Fuchs3384} M.~Fuchs, W.~G\"otze, M.~R. Mayr, 
    Phys.\ Rev.\ E {\bf 58}, 3384 (1998).
\bibitem{ref:Cohen1077} M.~H.~Cohen and G.~S.~Grest,
    Phys.\ Rev.\ B {\bf 20}, 1077 (1979).
\bibitem{ref:Bendler3970} J.~T.~Bendler and M.~F.~Shlesinger,
  J.\ Phys.\ Chem {\bf 96}, 3970 (1992).
\bibitem{ref:Chudley353-Dyre7243} 
  C.~T.~Chudley and R.~J.~Elliott,
  Proc.\ Phys.\ Soc. {\bf 77}, 353 (1961);
   J.~W.~Haus, K.~W.~Kehr, and J.~W.~Lyklema,
   Phys.\ Rev.\ B {\bf 25}, 2905 (1982); 
  R. Zwanzig, J.\ Chem.\ Phys. {\bf 79}, 4507 (1983);
  J. Machta and R.~Zwanzig,
   Phys.\ Rev.\ Lett. {\bf 50}, 1959 (1983);
   T. Odagaki and Y.~Hiwatari, Phys.\ Rev.\ A {\bf 41}, 929 (1990);
  T. Odagaki and Y.~Hiwatari, Phys.\ A {\bf 204}, 464 (1994);
  J.~C.~Dyre and J.~M.~Jacobsen, Phys.\ Rev.\ E {\bf 52}, 2429 (1995);
   C. Monthus and J.-P.~Bouchaud, J.\ Phys.\ A:Math.\ Gen. {\bf 29}, 3847 (1996); 
  J.~C.~Dyre and J.~M.~Jacobsen, Chem.\ Phys. {\bf 212}, 61 (1996); 
  A.~K.~Hartmann and D.~W.~Heermann, J.\ Chem.\ Phys. {\bf 108}, 9550 (1998);
   J.~C.~Dyre, Phys.\ Rev.\ E {\bf 59}, 2458 (1999);
  J.~C.~Dyre, Phys.\ Rev.\ E {\bf 59}, 7243 (1999).
\bibitem{ref:Dyre457-Schroeder3173} 
  J.~C.~Dyre, Phys.\ Lett. {\bf 108A}, 457 (1985);
   T. Odagaki, J.\ Phys.\ A:Math.~Gen. {\bf 20}, 6455 (1987);
  T. Odagaki, Phys.\ Rev.\ B {\bf 38}, 9044 (1988);
   J.~C.~Dyre, J.~Appl~Phys. {\bf 64}, 2456 (1988);
   J.~C.~Dyre and Th.~B.~Schr{\o}der, Phys.\ Rev.\ B {\bf 54}, 14884 (1996);
  Th.~B.~Schr{\o}der and J.~C.~Dyre, Phys.\ Rev.\ Lett. {\bf 84}, 310 (2000);
   J.~C.~Dyre and Th.~B.~Schr{\o}der, Phys.\ Stat.\ Sol.(b) {\bf 230}, 5 (2002);
   Th.~B.~Schr{\o}der and J.~C.~Dyre, Phys.\ Chem.\ Chem.\ Phys. {\bf 4}, 3173 (2002).
\bibitem{ref:Miyagawa3879} H. Miyagawa, Y. Hiwatari, B.~Berne, 
  and J.~P.~Hansen, J.\ Chem.\ Phys. {\bf 88}, 3879 (1988).
\bibitem{ref:kob_andersenI} W. Kob and H.C. Andersen, Phys.\ Rev.\ E
     {\bf 51}, 4626 (1995). 
\bibitem{ref:kob_andersenII} W. Kob and H.C. Andersen, 
  Phys.\ Rev.\ Lett. {\bf 73}, 1376 (1994); 
  Phys.\ Rev.\ E {\bf 52}, 4134 (1995).
\bibitem{ref:initconf}
  The 10 initial configurations differ drastically in their thermal
  history and are therefore completely independent. They were started at
  $T=5.0$ and, after various cooling and reheating periods, were
  equilibrated at $T=0.446$. Details can be found in 
  \cite{ref:kob_andersenI,ref:gleim98}.
\bibitem{ref:gleim98} T. Gleim and W. Kob, Phys. Rev. Lett. {\bf 81}, 
       4404 (1998).
%
\bibitem{ref:refnoteIII1} As described in Sec.~\ref{sec:model}, 
we start with the periodically (every 2000 MD steps of step size 
$\Delta t = 0.02$) stored
configurations
$\{{\bf r}_i(t)\}$ where $t=m \cdot 2000 \cdot 0.02$. For the single
particle fluctuations we use the time averages 
$\overline{{\bf r}_i}(m)= \frac{1}{20} \sum
\limits_{m'=m-9}^{m'=m+10} {\bf r}_i(m')$, their squares
$\overline{{\bf r}_i^2}(m)= \frac{1}{20} \sum
\limits_{m'=m-9}^{m'=m+10} {\bf r}_i^2(m')$,  and 
$\sigma_i^2(m) = \overline{{\bf r}_i^2}(m) 
          - \left ( \overline{{\bf r}_i}(m) \right )^2$ at 
times $m=10, 30, \ldots$. We first make a preliminary estimate of
the fluctuations 
$\sigma_{i,\ind{preest}}= \sqrt{\frac{1}{N_m}\sum \limits_{m} 
       \sigma^2_i(m)}$ 
and then obtain the final estimate for the fluctuations
$\sigma_{i,\ind{est}}$ by redoing the average over $\sigma^2_i$ but by 
averaging only over all
$\sigma^2_i(m)$ for which $\sigma_i^2(m) < 5 \sigma_{i,\ind{preest}}^2$
to roughly exclude jumps  from the average.
%
%
\bibitem{ref:refnoteIII3} We define ${\bf r}_i(t)$ to be a spike if 
$\left ( {\bf r}_i(t) - \langle \overline{{\bf r}_i} \rangle_\ind{f}
\right )^2 > 20 \langle \sigma_i^2 \rangle_\ind{f}$ 
where $\langle \cdot \rangle_\ind{f}$ is defined as in
Sec.~\ref{sec:jumplengths}. Similar to case 1 we say that the spike 
${\bf r}_i(t)$  returns to the previous average position if 
$\left ( {\bf r}_i(t) - \langle \overline{{\bf r}_i} \rangle_\ind{i}
\right )^2  \le 5 \langle \sigma_i^2 \rangle_\ind{i}$ .
%
\bibitem{ref:Oligschleger1031} C.~Oligschleger and H.~R.~Schober,
    Solid~State~Commun. {\bf 93}, 1031 (1995).
\bibitem{ref:Vardeman2568}  C.~F.~Vardeman~II and J.~D. Gezelter,
   J.\ Phys.\ Chem.\ A {\bf 105}, 2568 (2001). 
\bibitem{ref:eweeks_subdiff} E.\ R. Weeks and H.\ L. Swinney, Phys.\
Rev.\ E {\bf 57}, 4915 (1998).
\bibitem{ref:fogleman} M.\ A. Fogleman, M.\ J. Fawcett and T.\ H.
Solomon, Phys.\ Rev.\ E {\bf 63}, 020101(R).
\bibitem{ref:refnoteVI1} To determine the starting time $t_\ind{i}$ and 
ending time $t_\ind{f}$ of a jump we use an approach similar to our
definition for the occurrence of a jump (Eq.~(\ref{eq:jumpoccur})).
We compare the differences in 
averaged positions $\Delta \overline{{\bf r}_i}(t)$ and the
fluctuations $\sigma_{i,\ind{est}}$ (see \cite{ref:refnoteIII1}).
If a jump has been detected at time $t_\ind{i,detect}$ (because 
$|\Delta \overline{\bf {r}_i}|^2(t_\ind{i,detect}) > 20
\sigma^2_{i,\ind{est}}$) then the starting time $t_\ind{i}$ is the 
largest time $t$ for which 
$|\Delta \overline{{\bf r}_i}|^2(t) \le 5 \sigma_{i,\ind{est}}$
within the boundaries of $t_\ind{i,lowest} < t_\ind{i} < t_\ind{i,detect}$,
where in the case of a first jump $t_\ind{i,lowest}$ 
corresponds to the beginning of the production run and
in the case of the existence of a previous jump with
$t_\ind{f}^\ind{prev}$ then $t_\ind{i,lowest}=t_\ind{f}^\ind{prev}$.
Similarly after a jump has started it is detected to be 
finished at the earliest 
time $t_\ind{f,detect}$ for which 
$|\Delta \overline{\bf {r}_i}|^2(t_\ind{f,detect}) 
      \le 5 \sigma_{i,\ind{est}}$.
The finishing time of the jump $t_\ind{f}$ is then the maximum of 
$t_\ind{i,detect}$ and $t_\ind{f,detect}-3200$
where we subtract $3200$ ($=4 \cdot 20 \cdot 2000 \cdot 0.02$) because 
  $\Delta \overline{\bf r}_i = 
           \overline{\bf r}_i(t) - \overline{\bf r}_i(t-3200)$
(see Eq.~(\ref{eq:deltaoverlri}) and \cite{ref:refnoteIII1} ).
If the condition for $t_\ind{f,detect}$ never occurs, then 
$t_\ind{f}$ is equal to the length of the simulation run. To determine 
$\Delta t_\ind{b}$ in case 2 
of Fig.~\ref{fig:jumptype}, we use for the first spike the 
time between the previous $t_\ind{i}$ and the time of the 
spike and for successive spikes their time differences.
%
\bibitem{ref:intrinsic_structure} F.~H.~Stillinger and T.~A.~Weber,
   J.\ Chem.\ Phys. {\bf 80}, 4434 (1984);
   T.~A.~Weber and F.~H.~Stillinger, Phys.\ Rev.\ B {\bf 32}, 5402 (1985).
\bibitem{ref:refnoteVIII2} In Fig.~\ref{fig:sigEi_sigEinojump}a and 
 Fig.~\ref{fig:sigEi_sigEinojump}b we 
   use the time average windows $\langle \cdot \rangle_\ind{i,f}$.
     In Fig.~\ref{fig:sigEi_sigEinojump}a we then 
    average over jumps of a certain kind. In the case of a particle
with multiple jumps this means that the times after the first jump and
before the last jump are counted twice. In
Fig.~\ref{fig:sigEi_sigEinojump}b however we compare the behavior of {\em
particles} (instead of jump events) and therefore count these time
windows only once.
\end{references}
\end{document}